\documentclass{article}

\usepackage{a4wide}
\usepackage{amssymb}
\usepackage{color}
\usepackage{listings}
\usepackage[nointegrals]{wasysym}
\usepackage{MnSymbol}
\usepackage{authblk}
\usepackage{url}
\usepackage{breakcites}
\usepackage{hyperref}

\definecolor{myviolet}{rgb}{0.6,0.0,0.65}
\definecolor{myblue}{rgb}{0.1,0.0,0.8}
\definecolor{mygreen}{rgb}{0,0.4,0.2}
\definecolor{mycomment}{rgb}{0.7,0.2,0.2}
\definecolor{myred}{rgb}{0.8,0.0,0.0}
\definecolor{myorange}{rgb}{0.6,0.2,0.2}

\lstdefinelanguage{coq}[]{Caml}{
keywords=[1]{Section,Definition,Defined,CoInductive,Coercion,Inductive,Fixpoint,
  Parameter,Local,Module,Import,Record,Structure,Axiom,Lemma,Proposition,Theorem,Notation,
  Reserved,End,Proof,Goal,Qed,Variable,Variables,Hypothesis,Let,Program,Canonical,Check,Fail,Example},
keywordstyle=\color{myviolet}\ttfamily,
morekeywords=[2]{match,with,end,Set,Prop,Type,fun,of,let,in,struct,if,is,then,else,return},
keywordstyle=[2]\color{mygreen}\ttfamily,
morekeywords=[3]{move,refine},
keywordstyle=[3]\color{myblue}\ttfamily,
morekeywords=[3]{field,lra,rewrite},
keywordstyle=[4]\color{myred}\ttfamily,
morekeywords=[4]{by}
}

\def\pchoiceleft{\triangleleft}
\def\pchoiceright{\triangleright}
\def\us{\char`\_}
\def\coq{\textsc{Coq}}
\def\mathcomp{\textsc{MathComp}}
\def\analysis{\textsc{MathComp-Analysis}}
\def\finmap{\textsc{Finmap}}
\def\monae{\textsc{Monae}}
\def\newterm#1{{\sl #1}}
\def\pchoice#1#2#3{#1 \pchoiceleft #2 \pchoiceright #3}
\def\Pchoice#1#2#3{#1 :\mspace{-5.75mu}\pchoiceleft #2 \pchoiceright\mspace{-5.75mu}: #3}
\def\gcm{geometrically convex monad}
\def\GCM{Geometrically Convex Monad}
\def\choicetypecat{{\mathcal{C}}_C}
\def\convtypecat{{\mathcal{C}}_V}
\def\scslconvtypecat{{\mathcal{C}_S}}
\def\typecat{{\mathcal{C}_T}}
\def\image#1#2{#1{}{@}\large( #2 \large)}
\def\fapply#1#2{#1 \# \large( #2 \large)}
\def\nattranssymbol{\leadsto}
\def\nattrans#1#2{#1 \nattranssymbol #2}
\def\pdelta{P_\Delta}
\def\pdeltaright{\pdelta^{\textrm{right}}}
\def\pdeltaleft{\pdelta^{\textrm{left}}}
\def\convnsymbol{{\largetriangleleft\mspace{-13.75mu}\largetriangleright}}
\def\convn#1#2{\convnsymbol_{\!#1}\, #2}
\def\altsymbol{\Square}
\def\altop#1#2{#1 \,\altsymbol\, #2}

\def\bindsymbol{\gg\mspace{-5.75mu}=}
\def\bindop#1#2{#1 \bindsymbol #2}
\def\vcompsymbol{\cdot}

\def\hcompsymbol{\ast}

\def\lubop{\bigsqcup}
\def\mydef#1{\overset{\textrm{def}}{=}}

\usepackage[cmtip,all]{xy}
\newcommand{\longsquiggly}{\xymatrix@C=1.5em{{}\ar@{~>}[r]&{}}}

\usepackage{tikz}
\usetikzlibrary{calc,positioning} % above of, etc.
\usetikzlibrary{arrows,shapes}
\usetikzlibrary{decorations.pathmorphing,decorations.markings}
\usetikzlibrary{patterns}
\usetikzlibrary{cd}

\pgfdeclarelayer{background}
\pgfdeclarelayer{foreground}
\pgfsetlayers{background,main,foreground}

\title{A Trustful Monad for Axiomatic Reasoning with Probability and Nondeterminism}

\author[1]{Reynald  Affeldt}
\affil[1]{National Institute of Advanced Industrial Science and Technology,
        Cyber Physical Security Center, Japan}

\author[2]{Jacques Garrigue}
\affil[2]{Nagoya University,
        Graduate School of Mathematics, Japan}

\author[3]{David Nowak}
\affil[3]{CNRS \& Lille University,
        CRIStAL, France}

\author[4]{Takafumi Saikawa}
\affil[4]{Nagoya University,
        Graduate School of Mathematics, Japan}

\begin{document}

\date{}

\maketitle

\begin{abstract}
The algebraic properties of the combination of probabilistic choice
and nondeterministic choice have long been a research topic in program
semantics.
This paper explains a formalization in the Coq proof assistant of a
monad equipped with both choices: the geometrically convex monad.
This formalization has an immediate application: it provides a model
for a monad that implements a non-trivial interface which allows for
proofs by equational reasoning using probabilistic and
nondeterministic effects.
We explain the technical choices we made to go from the literature to
a complete Coq formalization, from which we identify reusable theories
about mathematical structures such as convex spaces and concrete
categories, and that we integrate in a framework for monadic
equational reasoning.
\end{abstract}

%\keywords{monad, formalization, probabilities, nondeterminism, convexity, categories}

\let\L=\lstinline

\section{Introduction}
\label{sec:intro}

In their ICFP paper ``Just {\tt do} It: Simple Monadic Equational
Reasoning''~\cite{gibbons2011icfp}, the authors present an axiomatic
approach to reason about programs with effects using equational
reasoning, thus recovering one of the appeals of pure functional
programming. This approach uses monads to encapsulate the effects,
hence the name \newterm{monadic equational reasoning}.
In particular, to handle the effects of probability and
nondeterminism, Gibbons and Hinze propose a combination of two
interfaces: one for monads equipped with an operator for probabilistic
choice and one for monads equipped with an operator for
nondeterministic choice. It was later observed that in the proposed
combination the authors ``got [the algebraic properties that
characterise their interaction] wrong''~\cite{abousaleh2016}.
The problem was that right-distributivity of bind over probabilistic
choice combined with distributivity of probabilistic choice
over nondeterministic choice resulted in an inconsistent theory.
Fortunately, the previous work in question~\cite{gibbons2011icfp} was
not relying on this mistake.

The example above shows that there is a need for a formal account of the
consistency of such a theory. One way to achieve it is to construct a monad
realizing the theory, which is, in our case, the combination of algebraic
theories of probabilistic and nondeterministic choices.
Monadic equational reasoning is not the only motivation to provide
a formalized monad.
Indeed, such a monad could be used to give semantics to programs
mixing probabilities and nondeterminism (e.g.,
\cite{kaminski2016esop}).
The infrastructure needed to formalize such a monad could be used to
formalize further foundational results in an area which is blooming
(e.g., \cite{bonchi2020presenting,mio2020monads,goy2020lics}).

In this paper, we provide a framework with which we formalize a monad
with an interface representing the combined algebraic theory of
probabilistic and nondeterministic choices; we moreover verify the
axiomatization of this theory and illustrate it with an example.
While many sets of axioms have been suggested as axiomatizations of
the combination of probabilistic and nondeterministic choice, only few
give rise to interesting models~\cite{mislove2004entcs,keimel2017lmcs}.
We will stick here to Gibbons and Hinze's axiomatization, removing
just the incriminated right-distributivity.
This gives us a trustful monad to reproduce Gibbons and Hinze's
examples of monadic equational reasoning.

We can rely on a large body of work to model formally the
combination of probabilistic and nondeterministic choice~(e.g.,
\cite{mislove2000concur,varacca2006mscs,beaulieu2008phd,tix2009entcs,gibbons12utp,keimel2017lmcs,cheungPhD2017},
and much more if we consider concurrency).
So what should be a monad modeling this axiomatization?
Since we already have the finite powerset monad and the
finitely-supported distributions monad for these two choices, one
could think of composing them.
At first sight, monadic distributive laws~\cite{beck1969} look like a
candidate approach but unfortunately it has been proved that
distributivity between these two monads is
impossible~\cite[Proposition~3.2]{varacca2006mscs}.
A very recent result~\cite{goy2020lics} indicates that weak
distributive laws provide a solution to this composition problem.
A more direct approach is to rethink the construction of a model of the
intended monad by looking into what it should be more precisely.
The presence of probabilistic choice suggests that sets of
distributions might be a model, like it is the case with the
probability monad.
Yet, the semantics must also be convex-closed because if two
distributions $d_1$ and $d_2$ are possible outcomes, so is any convex
combination $pd_1 + (1-p)d_2$ $(0\leq p \leq 1)$ of 
them~\cite[Sect.~5.2]{gibbons12utp}.
Convexity is in particular necessary to allow for idempotence of
probabilistic choice.
Unfortunately these observations do not readily lead to a
formalization, as they leave many technical details unsettled.
In his PhD thesis, Cheung derives a monad (called the
\newterm{\gcm{\/}}) for the theory resulting from the combination of
the effects of probability and
nondeterminism~\cite[Chapter~6]{cheungPhD2017}.
It highlights in particular the central role of \newterm{convex
  spaces}~\cite{stone1949,jacobs2010tcs,fritz2015arxiv} to formalize
convexity without resorting to vector spaces.

\paragraph*{Contributions}~%
In this paper, we provide a construction of the \gcm{} that can be
formalized by integrating reusable components (some obtained by
adapting existing work and some created for this occasion).
To the best of our knowledge, this is the first formalization of the
monad that combines probabilistic and nondeterminism choices
while retaining idempotence of probabilistic choice.
It has been carried out in the \coq{} proof assistant~\cite{coq}.
This construction is original; in particular, we adapt the
pencil-and-paper construction of Cheung~\cite{cheungPhD2017} to an
infinitary setting using Beaulieu's operator for infinite
nondeterministic choice~\cite[Def.~3.2.3]{beaulieu2008phd}.
We partly build on previous formalization work: theories of convex
spaces~\cite{saikawa2020cicm}, interfaces for monadic equational
reasoning, and finitely-supported probability
distributions~\cite{affeldt2019mpc}.
The new components that complete the construction of the \gcm{} are: a
formalization of the (non-empty) convex powerset functor~\cite[Sect.~5.1]{bonchi2017concur} and affine functions
(based on convex spaces), a formalization of semicomplete semilattice
structures (related to Beaulieu's work), and an original formalization
of concrete categories.  They are built in a reusable way following in
particular the methodology of packed
classes~\cite{garillot2009tphols}.
We will discuss how our choices allow these distinct formalizations to
fit together.
All these formal libraries are now available to tackle similar
formalizations that are already numerous as explained above.
Our formalization of the \gcm{} already has a direct application: it
is used to complete an existing formalization of monadic equational
reasoning called \monae{}~\cite{affeldt2019mpc}.
The latter comes with concrete monads modeling several interfaces {\em
  except\/} the one that combines probabilistic and nondeterministic
choices, because it is arguably more difficult than the others.
Our work improves the trusted base of this practical tool by filling
this hole.

\paragraph*{Paper Outline}~%
In Sect.~\ref{sec:target_approach}, we clarify our formalization
target by reviewing the formalization of monadic equational reasoning
we aim at extending.  We explain the operators of
interest and their properties, and we give an overview of the
construction of the \gcm{}.
In Sect.~\ref{sec:convexity}, we give an overview of a formalization
of convex spaces, an important ingredient of our construction to
represent probabilistic choice, convex sets, hulls, and affine functions.
In Sect.~\ref{sec:semicompsemilatt}, we explain the formalization of
semicomplete semilattice structures, which provide an operator to
represent a nondeterministic choice compatible with the probabilistic
choice.
In Sect.~\ref{sec:categories}, we explain a formalization of concrete
categories to build monads out of adjoint functors.
In Sect.~\ref{sec:adjoint_functors}, we define several adjunctions,
from which we derive the \gcm{} through composition.
In Sect.~\ref{sec:prop_comb_choice}, we verify that the
\gcm{} can be equipped with the combined choice and that the latter
enjoys the expected properties.
Finally, we show that the monad we have formalized can be used to
support monadic equational reasoning; we provide in
Sect.~\ref{sec:monty_hall} a {\em complete} mechanization of the Monty
Hall problem as presented by Gibbons.
We further comment on related work in Sect.~\ref{sec:related}
and conclude in Sect.~\ref{sec:conclusion}.

\paragraph*{About Notations}~%
For the sake of clarity, we try to display the \coq{} source code as
it is.
However, to limit the amount of code, we often indicate the
surrounding namespace using a comment instead of displaying the
precise \coq{} constructs (most of the time, this means that the name
of the surrounding \L!Module! appears as a comment for the reader to
figure out the fully qualified names).
To further ease reading, we perform some beautification using \LaTeX{}
symbols instead of ASCII art.
When there are too many details, we omit parts of the source code (and
mark them as ``\L!...!'') and instead provide a paraphrase and
indicate to the reader where to look in the formalization.
In the prose, we use as much as possible standard mathematical
notations, sometimes augmented to avoid too much overloading (for
example, we note $\image{f}{X}$ the direct image of the set~$X$ by~$f$
but $\fapply{F}{g}$ the application of the functor $F$ to the
morphism~$g$).

\paragraph*{About the Formalization}~%
This paper comes with a \coq{} formalization which is available online
as open source software~\cite{infotheo,monae}.

\section{Formalization Target and Approach}
\label{sec:target_approach}

Our goal is to construct a monad that combines probabilistic and
nondeterministic choices, as intended by
Gibbons et al.~\cite{gibbons2011icfp}. Here, we review an existing formalization
in \coq{} of Gibbons et al.'s monads and their interfaces~\cite{affeldt2019mpc}; our
formalization target is the model of the monad of type \L!altProbMonad!.

\subsection{An Existing Hierarchy of Probability-related Monads}
\label{sec:target}

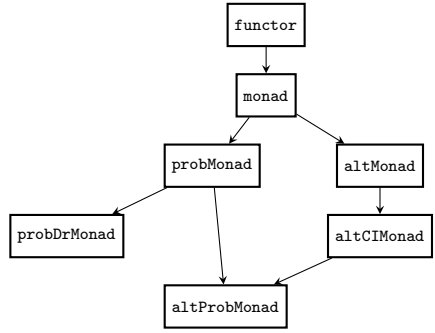
\begin{figure}[h]
\centering
\begin{tikzpicture}[node distance=10pt and 15pt,
      nodestyle/.style={scale=0.8,rectangle,minimum size=6, minimum height=20pt,thick,draw=black, font=\ttfamily\footnotesize},
      packedclass/.style={rectangle,fill=white},
      baseextension/.style={->,>=stealth,draw=black}]
  \node (altprobmonad) [nodestyle,packedclass] {altProbMonad};
  \node (altcimonad) [nodestyle,packedclass,above right=of altprobmonad] {altCIMonad};
  \path (altcimonad) edge[baseextension] (altprobmonad);
  \node (probdrmonad) [nodestyle,packedclass,above left=of altprobmonad] {probDrMonad};
  \node (probmonad) [nodestyle,packedclass,above right=of probdrmonad] {probMonad};
  \path (probmonad) edge[baseextension] (probdrmonad);
  \path (probmonad) edge[baseextension] (altprobmonad);
  \node (altmonad) [nodestyle,packedclass,above=of altcimonad] {altMonad};
  \path (altmonad) edge[baseextension] (altcimonad);
  \node (monad) [nodestyle,above left=of altmonad] {monad};
  \node (functor) [nodestyle,packedclass,above=of monad] {functor};
  \path (monad) edge[baseextension] (probmonad);
  \path (monad) edge[baseextension] (altmonad);
  \path (functor) edge[baseextension] (monad);
\end{tikzpicture}
\caption{Hierarchy of effects related to the monad type {\tt
    altProbMonad} that combines nondeterministic and probabilistic
  choices.}
\label{fig:hier}
\end{figure}

Figure~\ref{fig:hier} provides an excerpt of an existing hierarchy of
effects formalized~\cite{affeldt2019mpc} in \coq{} that includes the ones by Gibbons et
al.~\cite{gibbons2011icfp,gibbons12utp} (amended as suggested
by Abou-Saleh et al.~\cite{abousaleh2016}).
The complete hierarchy can be found in the online development~\cite[file \L!hierarchy.v!]{monae}.

\def\stt#1{{\small\texttt{#1}}}

We assume given two \newterm{types} \L!functor! and \L!monad! for
endofunctors and monads on Coq's \L!Type! universe. The type \L!monad!
is equipped with a join operator \L!Join! and a unit operator \L!Ret!.
In this section we rather use the bind operator, defined
as $\bindop{m}{f} \mydef \stt{Join}(\fapply{M}{f}\,m)$ for the monad~\L!M!.
The precise definitions of \L!functor! and \L!monad! are not relevant
at this stage but can be found in related
work~\cite[Sect.~2.1]{affeldt2019mpc}.

Note that these so-called ``types'' are actually data-structures that
provide the same functionality as type classes in Agda~\cite{agda} or
Idris~\cite{brady13jfp}, i.e., providing an implementation for such a
type amounts to defining an instance of the corresponding type
class. Moreover, thanks to implicit coercions, this implementation
itself can be used as a type, so that assuming \L!M : monad! allows
one to write the type \L!M T! of computations resulting in a value of
type \L!T! inside the monad~\L!M!.
The other nodes represent various monad types, that extend
\L!monad! through the incremental additions of \newterm{mixins}, using
the methodology of \newterm{packed classes}~\cite{garillot2009tphols}.

We first extend the type \L!monad! into the type of the probability
monad \L!probMonad!. The interface of \L!probMonad!
takes the form of a mixin that introduces an operator for
probabilistic choice $\pchoice{a}{p}{b}$, where $a$ and $b$ are
computations and $p$ is a \newterm{probability}, i.e., a real number
$p$ such that $0 \leq p \leq 1$.
The intuition is that the computation $\pchoice{a}{p}{b}$ represents
the computation $a$ with probability $p$ or the computation~$b$ with
probability $1-p$.
The properties, or \newterm{axioms}, of the interface are identity axioms
(lines~\ref{line:id0} and~\ref{line:id1}), skewed commutativity
(line~\ref{line:skewedcom}), idempotence
(line~\ref{line:idempotence}), quasi-associativity
(line~\ref{line:quasiassoc}), and the fact that bind left-distributes
over probabilistic choice.
\begin{lstlisting}[numbers=left,xleftmargin=2em,escapeinside=77]{coq}
(* Module MonadProb. *)
Record mixin_of (M : monad) : Type := Mixin {
  choice : forall (p : prob) T, M T -> M T -> M T
           where "a <| p |> b" := (choice p a b) ;
  _ : forall T (a b : M T), a <| 0%:pr |> b = b ; 7\label{line:id0}7
  _ : forall T (a b : M T), a <| 1%:pr |> b = a ; 7\label{line:id1}7
  _ : forall T p (a b : M T), a <| p |> b = b <| p.~%:pr |> a ; 7\label{line:skewedcom}7
  _ : forall T p (a : M T), a <| p |> a = a ; 7\label{line:idempotence}7
  _ : forall T (p q r s : prob) (a b c : M T), 7\label{line:quasiassoc}7
      (p = r * s :> R /\ s.~ = p.~ * q.~)%R ->
      a <| p |> (b <| q |> c) = (a <| r |> b) <| s |> c ;
  _ : forall p A B (m1 m2 : M A) (k : A -> M B), 7\label{line:leftdistbindpchoice}7
      (m1 <| p |> m2) >>= k = m1 >>= k <| p |> m2 >>= k}. 
\end{lstlisting}
In \coq{}, the type \L!prob! is for probabilities.
The notation \L!%:pr! turns a real number into a probability when
possible.
The notation $p$\L!.~! is for \L!1 - !$p$ (often written
$\overline{p}$ on paper).
Skewed commutativity allows to derive one of the identity axioms from
the other; here we are just preserving the original interface
from Gibbons and Hinze~\cite{gibbons2011icfp}.

The monad type \L!probDrMonad! extends \L!probMonad! with
right-distributivity of bind over probabilistic choice.  We do not
display its implementation because we do not model this monad in this
paper; we mention it for the sake of completeness.

The monad type \L!altMonad! introduces an operator \L![~]! for
nondeterministic choice\footnote{Gibbons and Hinze actually call
  ``choice'' and use the identifier {\tt\footnotesize alt} for what we call
  nondeterministic choice; they call nondeterministic choice a
  combination of choice and
  failure~\cite[Sect.~4.3]{gibbons2011icfp}.}. Besides associativity
of nondeterministic choice (line~\ref{line:altassoc} below), it also
states that bind left-distributes over nondeterministic choice
(line~\ref{line:leftdistbindalt}), as specified by the following mixin:
\begin{lstlisting}[numbers=left,xleftmargin=2em,escapeinside=77]{coq}
(* Module MonadAlt. *)
Record mixin_of (M : monad) : Type := Mixin {
  alt : forall T, M T -> M T -> M T where "a [~] b" := (alt a b) ;
  _ : forall T (x y z : M T), x [~] (y [~] z) = (x [~] y) [~] z ; 7\label{line:altassoc}7
  _ : forall A B (m1 m2 : M A) (k : A -> M B), 7\label{line:leftdistbindalt}7
      (m1 [~] m2) >>= k = m1 >>= k [~] m2 >>= k }.
\end{lstlisting}
Gibbons and Hinze do not require right-distributivity (i.e.,
$\bindop{m}{(\lambda x.\,\altop{k_1 x}{k_2 x})} =
\altop{(\bindop{m}{k_1})}{(\bindop{m}{k_2})}$) by default, due in
particular to undesirable interactions with non-idempotent
effects~\cite[Sect.~4.2]{gibbons2011icfp}.

The monad type \L!altCIMonad! extends \L!altMonad! with commutativity and
idempotence of nondeterministic choice, as expressed by the
following mixin, where \L!op x y! stands for \L!x [~] y!:
\begin{lstlisting}{coq}
(* Module MonadAltCI. *)
Record mixin_of (M : Type -> Type) (op : forall {T}, M T -> M T -> M T) : UU1 :=
  Mixin { _ : forall T (x : M T),   op x x = x ;
          _ : forall T (x y : M T), op x y = op y x }.
\end{lstlisting}

Finally, in the monad type \L!altProbMonad!, probabilistic choice
distributes over nondeterministic choice is expressed by
another mixin, where \L!op p x y! is intended to denote \L!x <| p |> y!:
\begin{lstlisting}{coq}
(* Module MonadAltProb. *)
Record mixin_of (M : altCIMonad) (op : prob -> forall {T}, M T -> M T -> M T) :=
  Mixin { _ : forall T p (x y z : M T), op p x (y [~] z) = op p x y [~] op p x z }.
\end{lstlisting}
%      x <| p |> (y [~] z) = (x <| p |> y) [~] (x <| p |> z) }.

\iffalse
Let us now discuss our choice of axioms in this last mixin.

We choose this distributivity axiom over the dual one, stating that
nondeterministic choice distributes over probabilistic choice, because
it has been argued that the latter does not match intuitions about
behavior, and happens to be inconsistent with idempotence.  Gibbons
provides an example of derivation that shows that probabilistic
choices become polluted with
nondeterminism with this axiom~\cite[Sect.~5.2]{gibbons12utp}.
%
This does not prevent some authors from adopting this opposite choice
(see the discussion by Gibbons~\cite[Sect.~9.1]{gibbons12utp}).

More importantly, one cannot have both distributivities, as it leads to
an almost total loss of meaning of the probability part of
probabilisitic choice: we have \L!x <| p |> y = x <| q |> y! for all
$p,q$ such that $0<p<q<1$~\cite[Thm.~A.3]{keimel2017lmcs}.

Another question is why we inherit from \L!probMonad! rather than
\L!probDrMonad!.
It appears that, while left-distributivity of bind over probabilistic
choice is fine, right-distributivity can be used to deduce the
distributivity of nondeterministic choice over probabilistic choice
from its dual, which is undesirable for the reason mentioned
above~\cite[Sect.~3]{abousaleh2016}.

Ultimately, the only way to be sure that our choice of axioms is
meaningful, is to provide a model where we can check that different
computations can be properly distinguished (which we will do in
Sect.~\ref{sec:nocollapse}).
\fi

\paragraph*{Implementation of Inheritance Relations with Packed Classes}~%
Up to now, we have only shown the mixin part of the inheritance
hierarchy. The packed class methodology~\cite{garillot2009tphols}
actually contains three ingredients: mixins, \newterm{classes}, and
\newterm{structures}. For example, here are the class and structure
definitions for \L!altProbMonad!.
\begin{lstlisting}[numbers=left,xleftmargin=2em,escapeinside=77]{coq}
(* Module MonadAltProb. *)
Record class_of (m : Type -> Type) := Class {
  base : MonadAltCI.class_of m ; 7\label{line:altciclass}7
  mixin_prob : MonadProb.mixin_of 7\label{line:probmixin}7
    (Monad.Pack (MonadAlt.base (MonadAltCI.base base))) ;
  mixin_altProb : @mixin_of 7\label{line:altprobmixin}7
    (MonadAltCI.Pack base) (@MonadProb.choice _ mixin_prob) }.
Structure altProbMonad : Type := Pack { 7\label{line:altprobstruct}7
  m :> Type -> Type ; class : class_of m }.
\end{lstlisting}
(In the code above, the modifier \L!@! disables implicit arguments and
the type declaration \L!:>! turns the corresponding structure field
into a coercion.)
The class definition inherits from \L!altCIMonad! through its class
(line~\ref{line:altciclass}), and extends it with two mixins: the one
we have seen for \L!probMonad!  (line~\ref{line:probmixin}) and the
additional distributivity axiom we have just defined
(line~\ref{line:altprobmixin}).  The structure
(line~\ref{line:altprobstruct}) then packages together the type
constructor~\L!m! with the class defined above.
Finally, the triple mixin-class-structure is completed with additional
coercions and unification hints (provided by the \L!Canonical!
command~\cite{mahboubi2013itp} of \coq{}) to achieve the inheritance
relations depicted in Fig.~\ref{fig:hier}.

\paragraph*{Sample Programs and Proof}~%
Last, for the sake of illustration, we reproduce sample programs by
Gibbons~\cite[Sect.~5.1]{gibbons12utp} and a simple proof by monadic
equational reasoning using the operators we have introduced so far in
the syntax of \monae{}. Here is a biased coin, with probability~$p$ of
returning \L!true! and probability $\overline{p}$ of returning
\L!false!:
\begin{lstlisting}{coq}
Definition bcoin {M : probMonad} (p : prob) : M bool :=
  Ret true <| p |> Ret false.
\end{lstlisting}
Here is an arbitrary nondeterministic choice between Booleans:
\begin{lstlisting}{coq}
Definition arb {M : altMonad} : M bool := Ret true [~] Ret false.
\end{lstlisting}
Using the {\tt do} notation instead of the bind operator, these two
programs can be used to make a probabilistic choice followed by a
an arbitrary choice:
\begin{lstlisting}{coq}
Definition coinarb p : M bool :=
  (do c <- bcoin p ; (do a <- arb; Ret (a == c) : M _))%Do.
\end{lstlisting}
Using \monae, one can prove that \L!coinarb! and \L!arb! are actually
the same program by means of mere rewritings:
\begin{lstlisting}{coq}
Lemma coinarb_spec p : coinarb p = arb.
Proof.
rewrite /coinarb /bcoin prob_bindDl !bindretf.
by rewrite /arb !alt_bindDl !bindretf eqxx eq_sym altC choicemm.
Qed.
\end{lstlisting}
Each lemma corresponds to an axiom from an interface.  In order,
\L!prod_bindDl! corresponds to left-distributivity of bind over
probabilistic choice, \L!bindretf! corresponds to the fact that
\L!Ret! is the left neutral of bind, \L!alt_bindDl!  corresponds to
left-distributivity of bind over nondeterministic choice, \L!altC!
corresponds to the commutativity of nondeterministic choice, and
\L!choicemm!  to the idempotence of probabilistic choice. Other
\L!rewrite! invocations are to unfold definitions or applying
properties of equality.
See Sect.~\ref{sec:monty_hall} for a larger example, and related work
for more proofs by monadic equational
reasoning~\cite{gibbons2011icfp,gibbons12utp,mu2019tr3,mu2019tr2} and
their mechanization~\cite{affeldt2019mpc}.

\subsection{Alternative Axiomatizations}
\label{sec:alternative}

As we mentioned in the introduction, the axiomatization of combined
choice we have followed is not the only possible one.
We will consider shortly two other possible axiomatizations for
which non-trival models are known.

The first one is obtained by replacing the distributivity axiom added
in \L!altProbMonad! by the dual one, i.e., distributivity of
nondeterministic choice over probabilistic choice:
\begin{center}
\L!x [~] (y <| p |> z) = (x <| p |> y) [~] (x <| p |> z)!.
\end{center}
Keimel et al.~\cite{keimel2017lmcs} have shown that in this case
probabilities different from~0 and~1 become indistinguishable (i.e.,
\L!x <| p |> y = x <| q |> y! for any $0 < p,q < 1$).
The algebraic theory of combined choice then boils
down to a bisemilattice (two semilattices with their operators
mutually distributing over each other).
This is equivalent to having both distributivity laws. While it
can be modeled by a powerset monad, the structure is poor, as
probability information is lost, so we did not try to formalize this
axiomatization.
Another way to reach the same axiomatization is to inherit from
\L!probDrMonad! rather than \L!probMonad!~\cite[Sect.~3]{abousaleh2016}.
It appears that, while left-distributivity of bind over probabilistic
choice is fine alone, right-distributivity can be used to deduce the
distributivity of nondeterministic choice over probabilistic choice
from its dual, which leads to the same collapse of probability
information as above.

The second one is obtained by keeping the same distributivity axiom as
in \L!altProdMonad!, but removing the idempotence of probabilistic
choice from \L!probMonad!, i.e., we lose the equality
\L!x <| p |> x = x!.
Varacca~\cite{varacca2006mscs} has shown that this relaxed
\L!probMonad! can be modeled by a monad of real quasi-cones, which
distributes over the finite powerset monad modeling \L!altCIMonad!.
As a result, one can use Beck's construction~\cite{beck1969} to create
a monad combining both. While this is a clever approach, the loss of
the idempotence axiom can be problematic depending on the application, and
Varacca presents in the same
paper another construction using a convex powerset functor to obtain a
model including the idempotence axiom, in a way similar to the
geometrically convex
powerdomain~\cite{mislove2000concur,tix2009entcs}.

Ultimately, the only way to be sure that our choice of axioms follows
our expectations, is to provide a model where we can check that different
computations can be properly distinguished (which we will do with the
\gcm{{} in Sect.~\ref{sec:nocollapse}).

\subsection{Formalization of the \GCM{}: Overview}
\label{sec:overview}

As already hinted at in the introduction (Sect.~\ref{sec:intro}), a
computation using the monadic operations defined in the type
\L!altProbMonad!  can be modeled by a non-empty convex set of
finitely-supported probability distributions.
Cheung provides a construction for such a monad and calls the
resulting monad the \gcm{}~\cite[Chapter~6]{cheungPhD2017}.  It is
built by composition of adjunctions, as depicted in Fig.~\ref{fig:gcm_cheung}.
The latter depicts three categories related by two adjunctions.
The category~{\sf Mod}({\sc prob}) corresponds to spaces with a convexity
operator (for probabilistic choice) and the category~{\sf Mod}({\sc
  prob}$\triangleright${\sc ndet}) corresponds to spaces with a convexity
operator and a binary operator (for nondeterministic choice).
The \gcm{} results from the composed adjunction
$F_1 \circ F_0 \dashv U_0 \circ U_1$.
We can derive the monad $U_0 \circ U_1 \circ F_1 \circ F_0$ directly from this adjunction~\cite{maclane1998}.

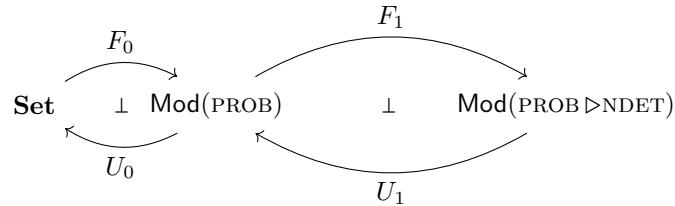
\begin{figure}[h]
\centering
\begin{tikzpicture}
\node (set) {\bf Set};
\node (modprob) [right=of set] {{\sf Mod}({\sc prob})};
\node [left=of modprob,xshift=6.5ex] {$\perp$};
\draw[->,transform canvas={yshift=0.5ex},bend left] (set) to node[above] {$F_0$} (modprob);
\draw[->,transform canvas={yshift=-0.5ex},bend left] (modprob) to node[below] {$U_0$} (set);
\node (modprobndet) [right=of modprob,xshift=7ex] {{\sf Mod}({\sc prob}$\triangleright${\sc ndet})};
\node [left=of modprobndet,xshift=3ex] {$\perp$};
\draw[->,transform canvas={yshift=0.5ex},bend left] (modprob) to node[above] {$F_1$} (modprobndet);
\draw[->,transform canvas={yshift=-0.5ex},bend left] (modprobndet) to node[below] {$U_1$} (modprob);
\end{tikzpicture}
\caption{Cheung's original diagram of adjunctions \cite[Fig.\ 6.1]{cheungPhD2017}}
\label{fig:gcm_cheung}
\end{figure}

Now that we have given an overview of the construction of the \gcm{},
let us take a step back to think ahead what we need to achieve its
formalization.
First, we need a formalization of convex spaces. This work has
actually started independently~\cite{saikawa2020cicm,affeldt2020cs}
and provides a formalization of convex spaces that can be easily
reused (among others, it develops a theory of convex functions).
Second, we need a formalization of probability distributions that can
be used as an instance of convex spaces and that can be used to form
the probability monad. Such a formalization happens to be available in
the form of a theory of finitely-supported probability
distributions~\cite{affeldt2019mpc}, which comes as an enhancement of
a theory of finite probability distributions~\cite{affeldt2014jar}
which could not be used to build a genuine monad because their type is
not an endofunction.
Third, we can draw inspiration from our previous work on formalizing
monadic equational reasoning~\cite{monae}. This work contains in
particular a formalization of the basic elements of Cheung's
construction (functors, adjunctions, monads, etc.) in the specialized
setting of the category {\bf Set}. Our experience with this work led
us more precisely to the following technical insights: (1)~packed
classes are a satisfactory approach to formalize the needed
mathematical structures, (2)~affine functions can be accommodated to
act as morphisms {\em provided\/} one uses concrete categories to
generalize from the specialized setting using the category {\bf Set}.
The very last bit of the story was to understand precisely the proofs
by Cheung's to realize that an infinitary operator for the
representation of the nondeterministic choice was called for
(namely, Beaulieu's operator already mentioned in
Sect.~\ref{sec:intro}).

We are now ready to recast Cheung's definition into the \coq{}
formalization we will explain in this paper.
Figure~\ref{fig:gcm} depicts four concrete categories related by three
adjunctions.
\begin{figure}[h]
\centering
\begin{tikzpicture}
\node (type) {\begin{tabular}{c} {\footnotesize\tt\color{mygreen}{Type}} \\ $\typecat$ \end{tabular}};
\node (choicetype) [right=of type,xshift=-1.6ex] {\begin{tabular}{c} {\footnotesize\tt choiceType} \\ $\choicetypecat$ \end{tabular}};
\node [left=of choicetype,xshift=6.5ex] {$\perp$};
\draw[->,transform canvas={yshift=0.5ex},bend left] (type) to node[above] {$F_C$} (choicetype);
\draw[->,transform canvas={yshift=-0.5ex},bend left] (choicetype) to node[below] {$U_C$} (type);
\node (convtype) [right=of choicetype,xshift=-1.6ex] {\begin{tabular}{c} {\footnotesize\tt convType} \\ $\convtypecat$ $$ \end{tabular}};
\node [left=of convtype,xshift=5ex] {$\perp$};
\draw[->,transform canvas={yshift=0.5ex},bend left] (choicetype) to node[above] {$F_0$} (convtype);
\draw[->,transform canvas={yshift=-0.5ex},bend left] (convtype) to node[below] {$U_0$} (choicetype);
\node (semicompsemilattconvtype) [right=of convtype,xshift=5.5ex] {\begin{tabular}{c} {\footnotesize\tt semiCompSemiLattConvType} \\ $\scslconvtypecat$ \end{tabular}};
\node [left=of semicompsemilattconvtype,xshift=5ex] {$\perp$};
\draw[->,transform canvas={yshift=0.5ex},bend left] (convtype) to node[above] {$F_1$} (semicompsemilattconvtype);
\draw[->,transform canvas={yshift=-0.5ex},bend left] (semicompsemilattconvtype) to node[below] {$U_1$} (convtype);
\end{tikzpicture}
\caption{Adjunctions between the categories involved in the construction of the \gcm{}}
\label{fig:gcm}
\end{figure}
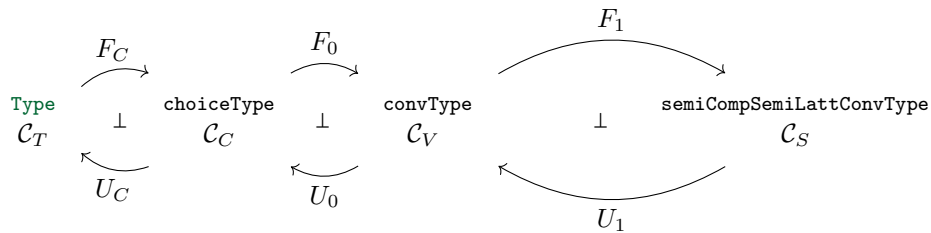
Each category is named after a \coq{} type to which it corresponds.
The category $\typecat$ corresponds to \coq{}'s type \L!Type!.  The
latter actually represents a countably infinite hierarchy of
types \L!Type!${}_0$, \L!Type!${}_1$, etc.\ such that \L!Type!${}_i$
is a subtype of \L!Type!${}_{i+1}$.  By default, \coq{}
hides the indices to the user.  We can regard \L!Type! as a category
by seeing each \L!Type!${}_i$ as a Grothendieck
universe~\cite{timany2016fscd}.
The
category~$\choicetypecat$ corresponds to types satisfying the axiom of
choice (i.e., equipped with a choice function).
The type \L!choiceType!~\cite[Sect.~3.1]{garillot2009tphols} comes
from the Mathematical Components library (hereafter,
\mathcomp{}~\cite{mathcomp}).
The category~$\convtypecat$ corresponds to {\sf Mod}({\sc prob}) and the
category~$\scslconvtypecat$ corresponds to a subcategory of {\sf Mod}({\sc
  prob}$\triangleright${\sc ndet}) with an infinitary operator for nondeterministic choice instead of a binary one; the
details of these two categories are one of the purposes of this paper.
The three adjunctions are composed of six functors. The unit
and counit of $F_C \dashv U_C$ are $\eta_C$ and $\varepsilon_C$
respectively (resp.\ $\eta_0$, $\varepsilon_0$ for $F_0 \dashv U_0$
and $\eta_1$, $\varepsilon_1$ for $F_1 \dashv U_1$). In particular,
$U_C$, $U_0$, and $U_1$ are forgetful functors, which makes $F_C$,
$F_0$, and $F_1$ free functors.
The desired monad $\pdelta = \pdeltaright \circ \pdeltaleft$ is obtained by  
composing adjunctions:
\[ \pdeltaleft = F_1 \circ F_0 \circ F_C \dashv U_C \circ U_0 \circ U_1 =
\pdeltaright. \]

Our setting features three adjunctions while Cheung's has only
two. The additional adjunction is the one between \L!Type!  and
\L!choiceType!. It comes from the fact that the formalization of
monadic equational reasoning we build upon~\cite{affeldt2019mpc}
represents monads as endofunctors over \L!Type!, whereas our
construction requires types to be equipped with a choice
function\footnote{Actually, these choice functions are not used in our
  development itself, but {\tt choiceType}s are required due to
  our use of the \finmap{} library~\cite{finmap} (which builds upon
  \mathcomp{}). See Sect.~\ref{sec:fcuc} for more details.}.  In practice, the functor $F_C$ only amounts to
adding a choice function to the type, without changing the values.
Note that, since we assume the existence of such a choice function for
all types, we are actually adding the axiom of choice to the ambient
logic, which is known to be sound in \coq~\cite{coqfaq}. It is simpler
to assume a well known axiom than to try to define all our monads on
\L!choiceType!, and prove that all the types we use can actually be
equipped with a concrete choice function.

\section{Convexity Toolbox}
\label{sec:convexity}

The formalization of the \gcm{} naturally calls for a formal theory of
convexity. As alluded to in Sect.~\ref{sec:overview}, it can be used
to represent the probabilistic choice, convex spaces (needed for the
categories~$\convtypecat$ and~$\scslconvtypecat$), non-empty convex
sets (to represent computations in a monad modeling \L!altProbMonad!),
convex hulls (to represent nondeterminism),
and also to represent the morphisms of the
categories~$\convtypecat$ and~$\scslconvtypecat$ (these morphisms are
affine functions) and the (non-empty) convex powerset functor~$F_1$.
For that purpose, we extend
an existing formalization of convex spaces~\cite{saikawa2020cicm}.

We recall our formalization of convex spaces in
Sect.~\ref{sec:convex_spaces}; its axiom system leads to a
formalization of convex sets and convex hulls, as explained in
Sect.~\ref{sec:convex_sets}.
We extend this formalization with affine functions and their
properties in Sect.~\ref{sec:affine_functions}.
The most relevant file of the online development for this
section is~\cite[file \L!convex_choice.v!]{infotheo}.

\subsection{Formalization of Convex Spaces}
\label{sec:convex_spaces}

A convex space (a.k.a.\ barycentric calculus~\cite{stone1949}) is an
algebraic structure allowing convex combinations of its elements by an
operator satisfying several equational axioms.
The interface is in fact similar to the interface of the \L!probMonad!
we saw in Sect.~\ref{sec:target}.
It provides an operator $\pchoice{a}{p}{b}$ where~$a$ and~$b$
are elements of the convex space and $p$~is a probability.
The axioms about the operator are similar to the ones already
explained in Sect.~\ref{sec:target} (the reader can observe a
difference of presentation for the axiom of quasi-associativity but it
is not relevant).
Of course, contrary to \L!probMonad!, convex spaces have no axiom
about a bind operator.
\begin{lstlisting}{coq}
(* Module ConvexSpace. *)
Record mixin_of (T : choiceType) : Type := Class {
  conv : prob -> T -> T -> T where "a <| p |> b" := (conv p a b);
  _ : forall a b, a <| 1%:pr|> b = a ;
  _ : forall p a, a <| p |> a = a ;
  _ : forall p a b, a <| p |> b = b <| p.~%:pr |> a;
  _ : forall (p q : prob) (a b c : T),
      a <| p |> (b <| q |> c) = (a <| [r_of p, q] |> b) <| [s_of p, q] |> c }.
\end{lstlisting}
The notation \L![s_of p, q]! stands for $\overline{\bar{p}\bar{q}}$;
the notation \L![r_of p, q]! stands for ${p}\,/\;{\overline{\bar{p}\bar{q}}}$.
Here we assume the carrier type of convex spaces to be a~\L!choiceType!.

The above mixin is used to define the type \L!convType! using the
packed classes methodology (that we briefly overviewed in Sect.~\ref{sec:target}).

We can show for example that the real numbers form a convex space by
taking the averaging function $\lambda p\,x\,y.\, px+\bar{p}y$ to be
the operator.
Similarly, finitely-supported probability distributions form a convex
space with the operator $\lambda p\,d_1\,d_2.\, pd_1+\bar{p}d_2$ where
$d_1$ and $d_2$ are distributions.

We will later need a generalization of the binary operator
$\pchoice{a}{p}{b}$ to $n$ points, namely $\convn{d}{f}$, where $f$
consists of $n$~points and $d$ is a distribution of $n$
probabilities.

\subsection{Convex Sets and Convex Hulls}
\label{sec:convex_sets}
%\label{sec:convex_hulls}

We use convex spaces to define \newterm{convex sets} and
\newterm{convex hulls}.
As already said in Sect.~\ref{sec:overview}, we put ourselves in a
classical setting that extends the logic of \coq{} with a number of
axioms known to be compatible with it.
Concretely, we use the axioms provided by \analysis{}, an extension of
\mathcomp{} for classical analysis~\cite{affeldt18jfr}.
In this setting, \L!Prop! and \L!bool! are equivalent (strong excluded middle),
and we can freely embed \L!Prop!-valued formulas such as \L!forall x, P x! into
\L!bool! using a notation: \L!`[<forall x, P x>] : bool!.
From \analysis{}, we also reuse a library of ``sets''.  Here ``sets''
means ``sets of elements of a specific type''. They are represented by
\L!Prop!-valued characteristic functions, and thus not necessarily
finite.  The type \L!set A! stands for sets over the type \L!A!.

A set~$D$ is convex when any convex combination of any two points is
still inside~$D$:
\begin{lstlisting}{coq}
Variable A : convType.
Definition is_convex_set (D : set A) : bool :=
  `[<forall x y t, D x -> D y -> D (x <| t |> y)>].
\end{lstlisting}

The hull of a set~$X$ is the set of points~$p$ such that $p$~is the
convex combination of points belonging to~$X$. The notation
\L![set p : T | P p]! is for sets defined by comprehension.
\begin{lstlisting}{coq}
Definition hull (T : convType) (X : set T) : set T :=
  [set p : T | exists n (g : 'I_n -> T) d, g @` setT `<=` X /\ p = <|>_d g].
\end{lstlisting}
We represent the $n$ points to be combined as $g_0, g_1, \ldots$,
hence the function \L!g : 'I_n -> T! from \L!'I_n!, the \mathcomp{}
type of natural numbers smaller than~\L!n!. The notation \L!g @` setT!
is for the direct image $\image{{\footnotesize\tt g}}{{\footnotesize\tt setT}}$ where
\L!setT!  is the full set (these are part of the library of sets that
comes with the \analysis{} library).

\subsection{Affine Functions}
\label{sec:affine_functions}

We are interested in \newterm{affine functions} because they are used
for the morphisms of the categories $\convtypecat$ and
$\scslconvtypecat$ (Sect.~\ref{sec:overview}). For example, in real
analysis, affine functions correspond to the functions of the form
$x \mapsto ax + b$. But the real line is just one example of convex
space. In fact, the generic operator of convex spaces provides an
easy, generic definition. First, we introduce a predicate that applies
to a function \L!f!, a pair of points \L!x! and \L!y!, and a
probability~\L!t!:
\begin{lstlisting}{coq}
Variables (T U : convType).
Definition affine_function_at (f : T -> U) x y t :=
  f (x <| t |> y) = f x <| t |> f y.
\end{lstlisting}
We use this predicate to characterize affine functions by a type for
\coq{} functions packaged with the following axiom:
\begin{lstlisting}
(* Module AffineFunction. *)
Variables (U V : convType).
Definition axiom (f : U -> V) := forall x y t, affine_function_at f x y t.
\end{lstlisting}
This packaging is done in such a way that affine functions can be used
as ordinary functions; see the online development~\cite{infotheo} for
details. Hereafter, the type of affine functions from the convex space
\L!U! to the convex space \L!V! is denoted by \L!{affine U -> V}!.

As a sample proposition, we can observe that convex hulls are
preserved by affine functions:
\begin{lstlisting}{coq}
Proposition image_preserves_convex_hull (f : {affine T -> U}) (Z : set T) :
  f @` (hull Z) = hull (f @` Z).
\end{lstlisting} 
This property will be used to define the functor $F_1$, whose action
on morphisms defined by the direct image needs to preserve convex
hulls.

\section{Semicomplete Semilattice Structures}
\label{sec:semicompsemilatt}

In this section, we define generic structures that provide an operator
to represent nondeterministic choice in a way that is compatible with
probabilistic choice.
The most relevant file in the online development is~\cite[file
\L!necset.v!]{infotheo}.

As a prerequisite, we introduce the type of non-empty sets.  The type
\L!neset T!  is the type of sets over~$T$ that have at least one
element. As a convenience, this type comes with a postfix notation
\L!%:ne! such that \L!s%:ne! is the non-empty set
corresponding to the set~\L!s!. This notation infers the proof of
non-emptiness in several situations such as when \L!s! is a singleton
set, the image of a non-empty set, the union of non-empty sets, etc.\
using \coq{}'s canonical structures~\cite{mahboubi2013itp}.

\subsection{Semicomplete Semilattice}
\label{sec:scsl}

The first structure we introduce provides a unary operator \L!op!  that
turns a non-empty set of elements into a single element
(line~\ref{line:scsl_op}).
The first axiom of this structure says that this operator applied to a
singleton set returns the sole element of the set
(line~\ref{line:scsl_ax1}).
The second axiom starting at line~\ref{line:scsl_ax2} collapses a
non-empty collection~\L!f! (the indexing set~\L!s! itself is not empty) of
non-empty sets into one element:
\begin{lstlisting}[numbers=left,xleftmargin=2em,escapeinside=77]{coq}
(* Module SemiCompleteSemiLattice. *)
Record mixin_of (T : choiceType) : Type := Mixin {
  op : neset T -> T ; 7\label{line:scsl_op}7
  _ : forall x : T, op [set x]%:ne = x ; 7\label{line:scsl_ax1}7
  _ : forall I (s : neset I) (f : I -> neset T), 7\label{line:scsl_ax2}7
        op (7$\displaystyle\bigcup_{\mbox{\tt i} \in \mbox{\small\tt s}}$7 f i)%:ne = op (op @` (f @` s))%:ne }.
\end{lstlisting}
The theory defined by this mixin is similar to Beaulieu's theory for
infinite nondeterministic choice~\cite[Def.~3.2.3]{beaulieu2008phd}.
The difference is that the right-hand side of the second axiom in
Beaulieu's work is expressed by means of a partition of the indexing
set. We prefer to avoid partitions because in our experience they cause
technical difficulties in formal proofs.

Hereafter we denote by $\lubop$ the operator introduced by the above
mixin and use the mixin to define the type \L!semiCompSemiLattType! of
\newterm{semicomplete semilattices}~\cite[p.~185]{bergman2015}.

\subsection{Combining Semicomplete Semilattice with Convex Space}
\label{sec:scslct}

We now extend the structure of semicomplete semilattices from the
previous section (Sect.~\ref{sec:scsl}) with an axiom that captures
the interaction between the operator $\lubop$  and probabilistic
choice. This interaction is akin to a distribution law that can be
stated informally as follows:
$$
\pchoice{x}{p}{\lubop\,I} = \lubop\,(\image{(\lambda\,y.\,\pchoice{x}{p}{y})}{I})
$$
Formally, this axiom is provided as a mixin parameterized by a
semicomplete semilattice and a ternary operator \L!op! indexed by a
probability:
\begin{lstlisting}{coq}
(* Module SemiCompSemiLattConvType. *)
Record mixin_of (L : semiCompSemiLattType) (op : prob -> L -> L -> L) :=
  Mixin { _ : forall (p : prob) (x : L) (I : neset L),
                 op p x (|_| I) = |_| ((op p x) @` I)%:ne }.
\end{lstlisting}
We use this mixin to extend the type of semicomplete semilattices to
the type of \newterm{semicomplete semilattice convex spaces}
(\L!semiCompSemiLattConvType! in \coq{} scripts) that inherits both
the properties of semicomplete semilattices (Sect.~\ref{sec:scsl}) and
the properties of convex spaces (Sect.~\ref{sec:convex_spaces}). The
methodology to achieve this multiple inheritance is again the one of packed
classes.

\def\oplusconvset{:\mspace{-5.75mu}\altsymbol\mspace{-5.75mu}:}

We conclude this section with a sample property of the operator
$\lubop$ that is both important and non-trivial:
\begin{lstlisting}{coq}
Variable L : semiCompSemiLattConvType.
Lemma lub_op_hull (X : neset L) : |_| (hull X)%:ne = |_| X.
\end{lstlisting}
The proof is as follows.
First, we lift the operator of convex spaces
$(\pchoice{}{p}{})$ from points to sets of points; we denote
this lifted operator by $(\Pchoice{}{p}{})$.
We use this lifted operator to define a new binary operator
$X \oplusconvset Y := \bigcup_{p \in [0,1]} \Pchoice{X}{p}{Y}$.
Second, we show that
$\displaystyle \texttt{hull}\,X =
\bigcup_{i\in\mathbb{N}}\underbrace{X \oplusconvset X \oplusconvset
  \cdots \oplusconvset X}_{i+1\textrm{ occurrences of }X}$.
Then, we show that
$\lubop(X) = \lubop(X \oplusconvset X \oplusconvset
\cdots \oplusconvset X)$, using the property introduced by
semicomplete semilattice convex spaces.
Finally, we conclude the proof by appealing to the properties of
semicomplete semilattices.

We will later provide a concrete example of use of the
lemma~\L!lub_op_hull!. It can also be used to establish technical
results
from Beaulieu's
work~(e.g., \cite[p.~56, l.~3]{beaulieu2008phd}) or
similar ones as in Varacca and Winskel's
work~(e.g., \cite[Lemma~5.6]{varacca2006mscs}).

\subsection{Instances with Non-empty Convex Sets}
\label{sec:scsl_instances}

The definitions of semicomplete semilattices and of semicomplete
semilattice convex spaces that we have provided in the previous
sections are just interfaces. To instantiate them, it turns out that
it suffices to use non-empty convex sets instead of mere non-empty
sets. This is this instance that we will use in particular to 
produce the adjunction $F_1 \dashv U_1$ (Fig.~\ref{fig:gcm}).

Thus we start by extending the type \L!neset! of non-empty sets into
the type \L!necset! of non-empty convex sets, using the definition
from the Sect.~\ref{sec:convex_sets} (and again the methodology of
packed classes).

We then instantiate the semicomplete semilattice operator on non-empty
convex sets using union and hull operators (\L!A! below is a convex
space):
$$
\begin{array}{lll}
\lubop &:&
\mbox{\L!neset (necset A) -> necset A!} \\
&&
X \mapsto \stt{hull}\left( \bigcup_{x\in X} x \right)
\end{array}
$$

This gives us in particular the type \L!necset_semiCompSemiLattConvType A!:
a generic instance of \L!semiCompSemiLattConvType! where the
carrier consists of non-empty convex sets over a \L!convType!~\L!A!.
We will use this type as the object part of
 $F_1 : \convtypecat \to \scslconvtypecat$.

The structures and instances explained in this section can be summarized
as the hierarchy pictured in Fig.~\ref{fig:hier_semicompsemilatt}.

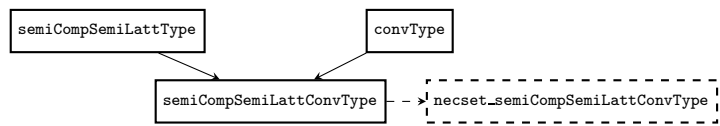
\begin{figure}[h]
\centering
\begin{tikzpicture}[node distance=10pt and 15pt,
      nodestyle/.style={scale=0.8,rectangle,minimum size=6, minimum height=20pt,thick,draw=black, font=\ttfamily\footnotesize},
      packedclass/.style={rectangle,fill=white},
      instance/.style={rectangle,dashed,fill=white},
      baseextension/.style={->,>=stealth,draw=black},
      instantiation/.style={->,>=stealth,dashed}]
  \node (semicompsemilattconv) [nodestyle,packedclass] {semiCompSemiLattConvType};
  \node (semicompsemilatt) [nodestyle,packedclass,above left=of semicompsemilattconv,xshift=1.5cm] {semiCompSemiLattType};
  \path (semicompsemilatt) edge[baseextension] (semicompsemilattconv);
  \node (conv) [nodestyle,packedclass,above right=of semicompsemilattconv,xshift=-1cm] {convType};
  \path (conv) edge[baseextension] (semicompsemilattconv);
  \node (necsetsemicompsemilattconv) [nodestyle,instance,right=of semicompsemilattconv] {necset\_semiCompSemiLattConvType};
  \path (semicompsemilattconv) edge[instantiation] (necsetsemicompsemilattconv);
\end{tikzpicture}
\caption{Hierarchy of semicomplete semilattices structures (dashed lines are for instances).}
\label{fig:hier_semicompsemilatt}
\end{figure}

\section{Formalization of Category Theory based on Concrete Categories}
\label{sec:categories}

The purpose of this section is to provide a formalization of enough
category theory to construct the \gcm{}. This formalization is
interesting in itself because it features an original use of concrete
categories through their shallow embedding.  It also fits our
application because it comes as a conservative extension of
\monae{}~\cite{affeldt2019mpc}.
The most relevant file in the accompanying development is~\cite[file
\L!category.v!]{monae}.

\subsection{Formalization based on Concrete Categories}

\subsubsection{Shallow Embedding of Concrete Categories}
\label{sec:cat_def}

As we saw in Sect.~\ref{sec:overview}, we need to formalize several
categories to formalize the \gcm; this is in contrast with \monae,
which could get along on the sole category of sets {\bf Set}.
Among the various possibilities, we chose to favor a definition akin
to a shallow embedding: it lets us use the typing relation of \coq{}
to declare elements of an object and apply morphisms to them as if
morphisms were ordinary \coq{} functions.
The starting idea is to represent categories with a \newterm{universe
  \`a la Tarski}, i.e., a type with an interpretation operation, or
\newterm{realizer}, allowing us to regard terms of this type as
\L!Type!s (the function~\L!el!  below at line~\ref{line:el}). In this
setting, we can then look at the morphisms of a category through the
realizer and identify the set of morphisms between two objects as a
subset of the function space between two realized objects (via the
predicate defining the hom-set at line~\ref{line:homset}).
\begin{lstlisting}[numbers=left,xleftmargin=2em,escapeinside=77]{coq}
(* Module Category. *)
Record mixin_of (obj : Type) : Type := Mixin {
  el : obj -> Type ; (* interpretation operation, "realizer" *) 7\label{line:el}7
  inhom : forall A B, (el A -> el B) -> Prop ; (* subset of morphisms *) 7\label{line:homset}7
  _ : forall A, @inhom A A idfun ; (* idfun is in inhom *)
  _ : forall A B C (f : el A -> el B) (g : el B -> el C),
      inhom f -> inhom g -> inhom (g \o f)
      (* inhom is closed by composition *) }.
Structure type : Type := Pack
  { carrier : Type ; class : mixin_of carrier }.
\end{lstlisting}
This definition has two salient features.
%
%First, we can use the subset relation defining hom-sets to consider
%subset of \coq{} functions, e.g., affine functions.
%
First, the parameter \L!obj! lets us choose how we index our objects
and use those indices to declare morphisms (e.g., \L!A! and \L!B! in
\L!f : el A -> el B!).
Second, we can use morphisms as functions and apply them to elements,
as illustrated by the following script:
\begin{lstlisting}{coq}
Variable C : category.
Variable A B : C.
Variable x : el A.
Variable f : {hom A, B}.
Check f x : el B.
\end{lstlisting}
Here, \L!{hom A, B}! is essentially the type of functions \L!el A -> el B!
equipped with a proof that they are morphisms (i.e., \L!f! such that
\L!inhom A B f! holds); there is a coercion from \L!{hom A, B}! to
\L!el A -> el B!.
Last, observe that, thanks to the shallow embedding, the laws of units
and composition are unnecessary because they are valid definitionally.

The resulting encoding is by no way ad hoc: it actually corresponds to
a shallow embedding of \newterm{concrete categories}.
A category $\mathcal C$ is said to be concrete if it comes with a faithful
functor from~$\mathcal C$ to~\textbf{Set}, that is, a functor whose action on
each hom-set is injective.
The indexing type \L!obj! and the realizer~\L!el! together form the
object part.
The function \L!inhom! represents the hom-sets of $\mathcal C$ by their images.
For the category-savy, the following diagram explains how the morphism
part~$F_\mathrm{Mor}$ of the faithful functor $F$ is represented through its
image in the hom-sets of {\bf Set}.
\newsavebox{\pullback}
\sbox\pullback{%
\begin{tikzpicture}%
\draw (0,0) -- (2ex,0ex);%
\draw (2ex,0ex) -- (2ex,2ex);%
\end{tikzpicture}}
%\begin{figure}[h]
%\centering
\begin{equation*}
\begin{tikzcd}
  &
  & \text{Im}({F_\mathrm{Mor}\restriction_{{\mathcal C}(A,B)}})
  \arrow[drr, phantom, "\usebox\pullback", very near start]
  \arrow[d, hook]
  \arrow[rr, "!"]
  &
  & * \arrow[d, "\text{\L!True!}"] \\
  {\mathcal C}(A,B)
  \arrow[rr, "{F_\mathrm{Mor}\restriction_{{\mathcal C}(A,B)}}"', tail]
  \arrow[rru]
  &
  & {\text{\bf Set}}(F(A),F(B)) \arrow[rr, "\text{\L!inhom A B!}"']
  &
  &  \text{\L!Prop!}
\end{tikzcd}
\end{equation*}
%\end{figure}
% 
Let $\mathcal C(A,B)$ be a hom-set of $\mathcal C$, which is mapped by
$F_\mathrm{Mor}$ (restricted to $\mathcal C(A,B)$) injectively into
the corresponding hom-set $\text{\bf Set}(F(A),F(B))$ of
$\text{\bf Set}$. Note that $\text{\bf Set}(F(A),F(B))$ appears in the
\coq{} code as the type \L!el A -> el B!.  The triangle on the left is
the image decomposition of
$F_\mathrm{Mor}\restriction_{{\mathcal C}(A,B)}$.  The square on the
right is a pullback diagram, with \L!inhom a b! being the
characteristic morphism of the image
$\mathrm{Im}({F_\mathrm{Mor}\restriction_{{\mathcal C}(A,B)}})$.

Except some hard examples (including homotopy categories), many
abstract categories can be concretized, i.e., we can find some
faithful functor from the category of {\bf Set} and rephrase it in our
framework.  The categories in this paper are concretized just by
injections, this is also the case for slice categories.  Other
examples require some encoding of objects and morphisms (e.g., product
categories).

\subsubsection{Categories to Build the \GCM{}}
\label{sec:cat_examples}

In this section, we instantiate our definition of concrete categories with the categories that were described in Sect.~\ref{sec:overview}.

\paragraph*{The Categories $\typecat$ and $\choicetypecat$}~%
We want to define the category $\typecat$, i.e., the situation in
which $\mathcal C$ is {\bf Set}, a.k.a.\ \L!Type!.  We just need to
instantiate $F_\mathrm{Mor}\restriction_{{\mathcal C}}$ to the
identity function and keep all the morphisms.
Technically, this amounts to instantiate the mixin
of the previous section with the identity function \L!fun x : Type => x!
as the realizer and the third argument of \L!@Category.Mixin! to be
the true predicate \L!fun _ _ _ => True!, so that the faithful functor
for the concrete category is full (i.e., surjective on hom-sets):
\begin{lstlisting}{coq}
Definition Type_category_mixin : Category.mixin_of Type :=
  @Category.Mixin Type (fun x : Type => x) (fun _ _ _ => True)
    (fun=> I) (fun _ _ _ _ _ _ _ => I).
Definition Type_category := Category.Pack Type_category_mixin.
\end{lstlisting}
(The identifier \L!I! is a proof of \L!True! in the standard library of \coq.)

Using this setting, we can now use the type \L!Type! of \coq{} as if
it were actually the category $\typecat$. The very last ingredient
is the declaration of \L!Type_category! as a canonical instance of
categories:
\begin{lstlisting}{coq}
Variable A : Type.
Fail Variable x : el A.
Canonical Type_category.
Variable x : el A.
\end{lstlisting}
The command \L!Canonical! (that we already mentioned for its use in
the packed classes methodology in Sect.~\ref{sec:target_approach})
provides a unification hint to \coq{}'s type-checker to automatically
endow \L!Type! with a structure of category when needed.
The other instances of categories in this section are also made canonical
but we only display the mixins which hold the relevant information.

Similarly to the category~$\typecat$, to define the category $\choicetypecat$, we take the function
\L!fun x : choiceType => Choice.sort x!, that returns the carrier
type (in \L!Type!) of its argument (we make \L!Choice.sort! appear explicitly
here but it is actually an implicit coercion in \coq{}). Again
the faithful functor is full:
\begin{lstlisting}{coq}
Definition choiceType_category_mixin : Category.mixin_of choiceType :=
  @Category.Mixin choiceType (fun x : choiceType => Choice.sort x)
    (fun _ _ _ => True) (fun=> I) (fun _ _ _ _ _ _ _ => I).
\end{lstlisting}

\paragraph*{The Category of Convex Spaces $\convtypecat$}~%
The objects are convex spaces (Sect.~\ref{sec:convex_spaces}) and the
morphisms are affine functions (between convex spaces), which can be
enforced by using the \L!axiom! from Sect.~\ref{sec:affine_functions}.
In our formalization, the objects are indexed by the type of convex spaces
\L!convType!, and realized by its coercion into \L!Type!.
Contrary to the previous two examples, being affine is not just a true predicate and requires us to prove that the identity function 
over a convex space is affine (proof \L!affine_function_id_proof!)
and that the composition of affine functions is affine
(proof \L!affine_function_comp_proof'!):
\begin{lstlisting}[escapeinside=77]{coq}
Definition convType_category_mixin : Category.mixin_of convType :=
  @Category.Mixin convType
    (fun A : convType => A) AffineFunction.axiom (* 7\color{myred}{Sect.~\ref{sec:affine_functions}}7 *)
    affine_function_id_proof affine_function_comp_proof'.
\end{lstlisting}

\paragraph*{The Category of Semicomplete Semilattice Convex Spaces $\scslconvtypecat$}~%
The objects are semicomplete semilattice convex spaces
(Sect.~\ref{sec:scslct}) and the morphisms are affine functions $f$
such that $\image{f}{\lubop\,X} = \lubop(\image{f}{X})$
for any non-empty convex set~$X$.
We can show that identity functions are such functions (proof
\L!lub_op_affine_id_proof!)  and that composition preserves these
properties (proof \L!lub_op_affine_comp_proof!), leading to the
following definition of $\scslconvtypecat$:
\begin{lstlisting}{coq}
Definition semiCompSemiLattConvType_category_mixin :
    Category.mixin_of semiCompSemiLattConvType :=
  @Category.Mixin semiCompSemiLattConvType
    (fun U : semiCompSemiLattConvType => U) LubOpAffine.class_of
    lub_op_affine_id_proof lub_op_affine_comp_proof.
\end{lstlisting}

\subsection{Formalization of Functors, Natural Transformations, and Monads}
\label{sec:fun_mon}

We now formalize functors, natural transformations, and monads using
the concrete categories formalized in the previous section.
In the following, \L!C! and \L!D! are two categories.

We encode a functor from \L!C! to \L!D! as an action
on objects represented by a function \L!m : C -> D!
(line~\ref{line:acto} below) and an action on morphisms represented by
a function
\L!f : forall A B,! \L!{hom A, B} ->! \L!{hom m A, m B}!
(line~\ref{line:actm}) equipped with proofs that \L!f! preserves the
identity (line~\ref{line:fid}) and composition (line~\ref{line:fo}):
\begin{lstlisting}[numbers=left,xleftmargin=2em,escapeinside=77]{coq}
(* Module Functor. *)
Record mixin_of (C D : category) (m : C -> D) : Type := Mixin { 7\label{line:acto}7
  f : forall (A B : Type), {hom A, B} -> {hom m A, m B} ; 7\label{line:actm}7
  _ : FunctorLaws.id f ; 7\label{line:fid}7
  _ : FunctorLaws.comp f }. 7\label{line:fo}7
\end{lstlisting}
By way of comparison, functors in \monae~\cite{affeldt2019mpc} were
specialized to the category {\bf Set} of sets and functions (the type
\L!Type! of \coq{} being interpreted as the category~{\bf Set}):
\begin{lstlisting}{coq}
Record mixin_of (m : Type -> Type) : Type := Class {
  f : forall (A B : Type), (A -> B) -> m A -> m B ;
  _ : FunctorLaws.id f ;
  _ : FunctorLaws.comp f }.
\end{lstlisting}
It is clear that the new, more general setting introduced above
improves on this specialized setting because it makes it possible to
talk about morphisms that are, e.g., affine functions. Hereafter,
we denote by \L!F # g! the application of a functor \L!F! to a morphism~\L!g!.

Let \L!F! and \L!G! be two functors from \L!C! to \L!D!. We encode a
natural transformation from \L!F! to \L!G! as a family 
of maps \L!f : forall A, {hom F A ,G A}! (hereafter,
denoted by \L!F ~~> G!) such the
\L!naturality! predicate holds:
\begin{lstlisting}{coq}
Variables (F G : functor C D).
Definition naturality (f : F ~~> G) := forall A B (h : {hom A, B}),
    (G # h) \o (f A) = (f B) \o (F # h).
\end{lstlisting}
When \L!F ~~> G! is packaged together with a proof of naturality,
we have a genuine natural transformation that we denote by~\L!F ~> G!
(mind the shorter arrow).

Finally, we define a monad as an endofunctor \L!M! equipped with two
natural transformations: \L!ret! from the identify functor (denoted by
\L!FId!) to \L!M!, and \L!join! from the composition of \L!M! with
itself (denoted by \L!M \O M!) to \L!M!. The proofs of naturality
appear at lines~\ref{line:nat_ret} and~\ref{line:nat_join}. These two
natural transformations furthermore satisfy three coherence conditions
(lines~\ref{line:left_ret_join}, \ref{line:right_ret_join},
and~\ref{line:joinA}):
\begin{lstlisting}[numbers=left,xleftmargin=2em,escapeinside=77]{coq}
(* Module Monad. *)
Record mixin_of (C : category) (M : functor C C) : Type := Mixin {
  ret : forall A, {hom A, M A} ;
  join : forall A, {hom M (M A), M A} ;
  _ : naturality FId M ret ; 7\label{line:nat_ret}7
  _ : naturality (M \O M) M join ; 7\label{line:nat_join}7
  _ : forall A, join A \o ret (M A) = id ; 7\label{line:left_ret_join}7
  _ : forall A, join A \o M # ret A = id ; 7\label{line:right_ret_join}7
  _ : forall A, join A \o M # join A = join A \o join (M A) }. 7\label{line:joinA}7
\end{lstlisting}
We already said above that our formalization of functors generalizes
the one of \monae{}, the formal framework for monadic equational
reasoning on which our work is based.  Our formalization of monads
also generalizes the one of \monae{} in a conservative way.
Concretely, we provide a function \L!Monad_of_category_monad! that
given a monad (as defined just above) over the category~$\typecat$,
returns a monad as defined in \monae{} (over \L!Type!, regarded as the
category {\bf Set}). This way, it will be possible to (1)~prove that
our formalization of the \gcm{} satisfies the expected axioms and
(2)~retrofit it back into \monae{}.

\subsection{Formalization of Adjoint Functors}
\label{sec:adjfun}

We use adjoint functors to build the \gcm{}. In this
section, we recall the lemmas used for this construction and give a
brief overview of their formalization. We do not provide all the
technical details because these lemmas are well-known lemmas and their
formalization follows naturally from the definitions we saw so far.

\subsubsection{Definition of Adjunction}
\label{sec:adjfun_def}

Two functors $F$ and $G$ are \newterm{adjoint\/} (denoted by
$F \dashv G$) when there are two natural transformations
$\eta : \nattrans{1}{G \circ F}$ and
$\varepsilon : \nattrans{F \circ G}{1}$ such that $\eta$ and
$\varepsilon$ satisfy the triangular laws
$\forall c.\, \varepsilon(F\,c) \circ \fapply{F}{\eta\,c}=id$
(triangular left) and
$\forall d.\, \fapply{G}{\varepsilon\,d} \circ \eta(G\,d) =id$
(triangular right).

In \coq{}, we provide the notation \L!F -| G! for the following type
(where the categories \L!C! and \L!D! are implicit arguments):
\begin{lstlisting}{coq}
AdjointFunctors.t : forall C D : category, functor C D -> functor D C -> Type
\end{lstlisting}
To build an adjunction, one needs to provide two natural
transformations \L!eta! and \L!eps! together with the proofs that they
satisfy the triangular laws. The corresponding constructor has
the following type (where all arguments except the proofs of the triangular laws
are implicit):
\begin{lstlisting}{coq}
AdjointFunctors.mk : forall (C D : category) (F : functor C D) (G : functor D C)
  (eta : FId ~> G \O F) (eps : F \O G ~> FId),
  TriangularLaws.left eta eps -> TriangularLaws.right eta eps -> F -| G
\end{lstlisting}

\subsubsection{Composition of Adjunction}
\label{sec:adjcomp}

It is well-known that two adjunctions $F \dashv G$ (with
unit/counit $\eta$/$\varepsilon$) and $F' \dashv G'$ (with unit/counit
$\eta'$/$\varepsilon'$) can be composed to form another
adjunction $F' \circ F \dashv G \circ G'$ by taking the unit to
be $\lambda A.\, \fapply{G}{\eta' ({F}_A)} \circ {\eta}_A)$ and the
counit to be
$\lambda A.\, \varepsilon'_A \circ \fapply{F'}{\varepsilon(G'_A)}$.
Using the constructs we have defined so far, we provide a \coq{}
function that performs this composition:
\begin{lstlisting}{coq}
adj_comp : forall (C0 C1 C2 : category)
  (F : functor C0 C1)  (G : functor C1 C0),  F -| G ->
forall (F' : functor C1 C2) (G' : functor C2 C1), F' -| G' ->
  F' \O F -| G \O G'
\end{lstlisting}

\subsubsection{Monad Defined by Adjointness}
\label{sec:monadofadjoint}

It is well-known that an adjunction $F \dashv G$ gives
rise to a monad $G \circ F$ by taking $\eta$ to be the unit and
$\lambda A.\,\fapply{G}{\varepsilon(F_A)}$ to be the join operator.
In our formalization, this construction takes the form of the following
function:
\begin{lstlisting}{coq}
Monad_of_adjoint : forall (C D : category) (F : functor C D) (G : functor D C),
  F -| G -> monad C
\end{lstlisting}
Observe that contrary to \monae{} where all monads are over the
category {\bf Set}, here our monad is over some category~\L!C! which
appears explicitly in the type.

\section{Adjoint Functors for the \GCM{}}
\label{sec:adjoint_functors}

At this point, we have explained the formalization of all the elements
necessary to construct the \gcm{}: convex spaces and affine functions in
Sect.~\ref{sec:convexity}, semicomplete semilattice structures in
Sect.~\ref{sec:semicompsemilatt}, and category theory in
Sect.~\ref{sec:categories}.
In this section, we explain the formalization of the adjunctions
explained in Sect.~\ref{sec:overview}.
The most relevant file from the accompanying development is~\cite[file
\L!gcm_model.v!]{monae}.

\subsection{The Adjunction $F_C \dashv U_C$}
\label{sec:fcuc}

The raison d'\^etre of the adjunction $F_C \dashv U_C$ in our
formalization is essentially technical: it comes from the use of the \coq{} type
\L!Type! in \monae{} and the need to use a \L!choiceType!  in the
definition of finitely-supported distributions.
% (this need itself is
% inherited from our use of the \finmap{} library~\cite{finmap} of
% \mathcomp{})

Let us first define the functor $F_C$ from $\typecat$
to~$\choicetypecat$.
The action on objects consists in turning a type in \L!Type!  into a
\L!choiceType!. This is performed by the function \L!choice_of_Type!
which relies on an axiom inherited from a \mathcomp{} library and
whose validity is explained elsewhere~\cite[Sect.~5.2]{affeldt18jfr}.
The action on morphisms turns a morphism $f : T \to U$ into the same
morphism but with type
\L!choice_of_Type!$\,T\to\,$\L!choice_of_Type!$\,U$\/:
\begin{lstlisting}{coq}
Definition hom_choiceType (A B : choiceType) (f : A -> B) : {hom A, B} :=
  HomPack (I : InHom (f : el A -> el B)).
Local Notation CT := Type_category.
Definition free_choiceType_mor (T U : CT) (f : {hom T, U}) :
  {hom m T, m U} := hom_choiceType (f : m T -> m U).
\end{lstlisting}
The purpose of the function \L!hom_choiceType! is to turn a \coq{}
function between two \L!choiceType!s into a morphism of the
category~$\choicetypecat$. Here, \L!I! (that we already saw in Sect.~\ref{sec:cat_examples})
acts as a trivial proof that \L!f! is indeed a morphism; it is sufficient
because in this category all
functions are morphisms (the notation \L!HomPack! is
just a smart constructor~\cite{monae}).
The functor laws are trivially proved and together with the
definitions above, this leads to the definition of the functor
\L!free_choiceType! of type \L!functor CT CC!.

The definition of the corresponding forgetful functor $U_C$ is
similar. The main difference is that instead of using the function
\L!choice_of_Type! to augment a type in \L!Type!, we use the coercion
\L!Choice.sort! that retrieves the carrier type of a \L!choiceType!
(see \L!forget_choiceType! in \cite[file \L!gcm_model.v!]{monae}).

\label{sec:counitunitC}

The unit $\eta_C : \nattrans{1}{U_C \circ F_C}$ and the counit
$\varepsilon_c : \nattrans{F_C \circ U_C}{1}$ are also essentially
identity functions and the proofs of the triangular laws are therefore
trivial.

\subsection{The Adjunction $F_0 \dashv U_0$}
\label{sec:f0u0}

The second adjunction $F_0 \dashv U_0$ corresponds to
the probability monad~\cite{giry1982}.
It relies on an existing formalization of finitely-supported
distributions~\cite[Sect.~6.2]{affeldt2019mpc} that we recall briefly.
In the definition of \L!FSDist.t! below, the first field
(line~\ref{line:fsdistf}) is a finitely-supported function~\L!f! from
the \L!choiceType! \L!A! to the type of real numbers from the standard
\coq{} library; this function evaluates to \L!0! outside its support
\L!finsupp f!.
The second (anonymous) field (line~\ref{line:fsdistP}) contains proofs that (1) the probability
function outputs positive reals and that (2) its outputs sum to~$1$.
\begin{lstlisting}[numbers=left,xleftmargin=2em,escapeinside=77]{coq}
(* Module FSDist. *)
Record t := mk {
  f :> {fsfun A -> R with 0} ; 7\label{line:fsdistf}7
  _ : all (fun x => 0 <b f x) (finsupp f) && \sum_(a <- finsupp f) f a == 1 } 7\label{line:fsdistP}7.
\end{lstlisting}
It is important to observe that \L!FSDist.t! has type
\L!choiceType -> choiceType! and can therefore be used to
build an endofunctor and a
monad on top of it.
Hereafter, \L!{dist A}! is a notation for \L!FSDist.t A!.

\subsubsection{Functors}

The action on morphisms of $F_0$ is the map
of the probability monad associated with
finitely-supported distributions. Indeed, let
$\pchoice{\cdot}{\cdot}{\cdot}$ be the operation of the convex space
of finitely-supported distributions (see
Sect.~\ref{sec:convex_spaces}) and let $\bindsymbol$ be the bind
operator of the probability monad. We have
$\bindop{(\pchoice{d_1}{p}{d_2})}{f} =
\pchoice{(\bindop{d_1}{f})}{p}{(\bindop{d_2}{f})}$, which is
equivalent to the map of the probability monad being affine.

In \coq{}, we define the action on morphisms of $F_0$ as follows,
where \L!FSDistfmap! is the map operation of the probability monad:
\begin{lstlisting}{coq}
Definition free_convType_mor (A B : choiceType) (f : {hom A, B}) :
  {hom FSDist_convType A, FSDist_convType B} :=
  @Hom.Pack CV _ _ _ (FSDistfmap f) (FSDistfmap_affine f).
\end{lstlisting}
The type \L!FSDist_convType A! is the type of convex spaces of
finitely-supported distributions over~\L!A! and \L!FSDistfmap_affine!
is the proof that \L!FSDistfmap f! is affine.

We can show that \L!free_convType_mor! satisfies the functor laws (proofs
\L!free_convType_mor_id! and \L!free_convType_mor_comp!), leading to
the definition of the functor $F_0$ (recall the definitions of
Sect.~\ref{sec:fun_mon}):
\begin{lstlisting}{coq}
Definition free_convType : functor CC CV :=
  Functor.Pack (Functor.Mixin free_convType_mor_id free_convType_mor_comp).
\end{lstlisting}
The constructors \L!Functor.Mixin! and \L!Functor.Pack! are
respectively for the mixin and the type of functors explained in
Sect.~\ref{sec:fun_mon}.

The forgetful functor $U_0$ of type \L!functor CV CC!  is just
formalized by substituting the category~\L!CV!  by the
category~\L!CC! in morphisms (see \L!forget_convType! in \cite[file
\L!gcm_model.v!]{monae}).

\subsubsection{Counit / unit}
\label{sec:counitunit0}

\def\finsupp#1{\stt{finsupp}\left( #1 \right)}

The counit is the natural transformation
$\varepsilon_0 : \nattrans{F_0 \circ U_0}{1_{\convtypecat}}$
essentially defined by the following function:
$$
\begin{array}{rl}
\varepsilon_0 : & \{\stt{dist}\,C\} \to C \\
& d \mapsto \convn{d}{\finsupp{d}}.
\end{array}
$$
In this definition, $C$ is a \L!convType!; the operation
``$\convn{\cdot}{\cdot}$'' has been explained in Sect.~\ref{sec:convex_spaces}.
Intuitively, $\varepsilon_0$ corresponds to the computation of a
barycenter.

The unit is the natural transformation 
$\eta_0 : \nattrans{1_{\choicetypecat}}{U_0 \circ F_0}$
defined by the point-supported distribution \L!FSDist1.d!:
$$
\begin{array}{rl}
\eta_0 : & C \to \{\stt{dist}\,C\} \\
& x \mapsto \stt{FSDist1.d}\;\;x.
\end{array}
$$

The proofs of the triangular laws required us to substantially enrich
the theory of finitely-supported distributions used in \monae.
The reason can be understood by looking at the proof of the left
triangular law \L!triL0!.
The latter essentially amounts to prove that we have for any
probability distribution~$d$\/:
$$
\convn{\stt{FSDistfmap}\;\stt{FSDist1.d}\;d}{\finsupp{\stt{FSDistfmap}\;\stt{FSDist1.d}\;d}}
= d.
$$
One can observe that this statement involves distributions of distributions
\begin{lstlisting}{coq}
Check FSDistfmap (@FSDist1.d C) d : {dist {dist C}}.
\end{lstlisting}
whose properties called for new lemmas.
Comparatively, the proof of the right triangular law \L!triR0! is simpler.

\subsection{The Adjunction $F_1 \dashv U_1$}
The third adjunction $F_1 \dashv U_1$ corresponds to the nondeterminism part of the
\gcm{}, giving a nondeterminism monad over the category $\convtypecat$ of convex
spaces.  It consists of the (non-empty) convex powerset functor $F_1$ and a corresponding
forgetful functor $U_1$.

\subsubsection{Functors}
\label{sec:adjunction1_functors}

The action on objects of $F_1$ is \L!necset_semiCompSemiLattConvType!,
explained in Sect.~\ref{sec:scsl_instances}.
The action on morphisms of $F_1$ is defined by the direct image
$f \mapsto \lambda X.\, \image{f}{X}$ (where $X$ is a non-empty convex set):
\begin{lstlisting}[escapeinside=77]{coq}
Variables (A B : convType) (f : {hom A, B}).
Definition free_semiCompSemiLattConvType_mor'
    (X : necset_convType A) : necset_convType B :=
  7NECSet.Pack7 (* definition using the direct image omitted *).
\end{lstlisting}
We can show that the image of a morphism is still a morphism: it is
affine and preserves $\lubop$ (because convex hulls are preserved by
taking the direct image along affine functions---Sect.~\ref{sec:affine_functions}):
\begin{lstlisting}{coq}
Definition free_semiCompSemiLattConvType_mor :
  {hom necset_semiCompSemiLattConvType A,
       necset_semiCompSemiLattConvType B} :=
  @Hom.Pack CS _ _ _ free_semiCompSemiLattConvType_mor'
    (LubOpAffine.Class free_semiCompSemiLattConvType_mor'_affine
                       free_semiCompSemiLattConvType_mor'_lub_op_morph).
\end{lstlisting}
To be more precise, this is the lemma
\L!free_semiCompSemiLattConvType_mor'_lub_op_morph!  that uses the
lemma \L!image_preserves_convex_hull! explained in
Sect.~\ref{sec:affine_functions}.

Finally, we show that the action on morphisms satisfies the functor
laws, leading to the following definition of $F_1$:
\begin{lstlisting}{coq}
Definition free_semiCompSemiLattConvType : functor CV CS :=
  Functor.Pack (Functor.Mixin free_semiCompSemiLattConvType_mor_id
                              free_semiCompSemiLattConvType_mor_comp).
\end{lstlisting}

Like for the adjunction $F_0 \dashv U_0$, the forgetful functor $U_1$ of
type \L!functor CS CV!  is just formalized by substituting the
category~\L!CS! by the category~\L!CV! in morphisms (see
\L!forget_semiCompSemiLattConvType! in \cite[file
\L!gcm_model.v!]{monae}).

\subsubsection{Counit / unit}
\label{sec:counitunit1}

Let us explain how we implement the counit
$\varepsilon_1 : \nattrans{F_1 \circ U_1}{1_\scslconvtypecat}$.

It is exactly the $\lubop$ operator seen in
Sect.~\ref{sec:scsl_instances}:
$$
\begin{array}{rl}
\varepsilon_1 : & \stt{neset}(\stt{necset}\,T) \to \stt{necset}\,T \\
& X \mapsto \lubop\,X.
\end{array}
$$
We need to show that it is natural, that it preserves the operator $\lubop$, i.e.,
$\varepsilon_1(\lubop(X)) =
\lubop(\image{\varepsilon_1}{X})$ (for that purpose we use the lemma \L!lub_op_hull! from
Sect.~\ref{sec:scslct}), and that it is affine, i.e.,
$\varepsilon_1(\pchoice{X}{p}{Y})=\pchoice{\varepsilon_1\,X}{p}{\varepsilon_1\,Y}$.

Let us comment on the proof that $\varepsilon_1$ preserves the
nondeterministic choice to highlight a key difference with Cheung's
work~\cite{cheungPhD2017}.
From the proof that $\varepsilon_1$ preserves the infinitary
nondeterministic choice~\cite[lemma \L!eps1''_lub_op_morph!, file
\L!gcm_model.v!]{monae}, we can derive the proof that it preserves the binary
nondeterministic choice~\cite[lemma \L!lub_op_lub_binary_morph!, file
\L!necset.v!]{infotheo}.
In contrast, Cheung proves the binary version directly.
Cheung's setting is finitary but his proofs rely on an implicit
connection between finitary and infinitary uses of convex hulls
which make them incomplete (at best).
This manifests concretely by the use of an undefined infinitary
operator~\cite[p.~160]{cheungPhD2017}.
We think that there is a way to make sense of his proof,
seeing it as using finitary operators on finite sets whose convex
hulls correspond to the infinite sets appearing in his proof, but the
theory underlying that reading is completely omitted.
Anyway, we have experienced that an infinitary setting is more
comfortable for formal proofs. Those are the main reasons why we think
that formalization is best performed in an infinitary setting.

The unit $\eta_1 : \nattrans{1_{\convtypecat}}{U_1 \circ F_1}$ is the
singleton map, which is easily shown to be natural and affine.
$$
\begin{array}{rl}
\eta_1 : & \stt{necset}\,T \to \stt{neset}(\stt{necset}\,T) \\
& X \mapsto \{T\}
\end{array}
$$

We call the corresponding triangular laws \L!triL1! and \L!triR1!.

\subsection{Putting it All Together}

\subsubsection{Formalization of the \GCM{}}
\label{sec:formal_gcm}

We use the proofs of the triangular laws of Sections
\ref{sec:counitunitC}, \ref{sec:counitunit0},
and~\ref{sec:counitunit1} to create the three adjunctions
$F_C \dashv U_C$, $F_0 \dashv U_0$, and $F_1 \dashv U_1$:
\begin{lstlisting}{coq}
Definition AC := AdjointFunctors.mk triLC triRC.
Definition A0 := AdjointFunctors.mk triL0 triR0.
Definition A1 := AdjointFunctors.mk triL1 triR1.
\end{lstlisting}
The definition of these adjunctions has been given in
Sect.~\ref{sec:adjfun_def}.

We then build the adjunction resulting from the
composition of the three adjunctions we have just defined, using the
function of Sect.~\ref{sec:adjcomp}:
\begin{lstlisting}{coq}
Definition Agcm := adj_comp AC (adj_comp A0 A1).
\end{lstlisting}

Finally, we obtain the \gcm{} from the resulting adjunction
using the generic lemma explained at the end of
Sect.~\ref{sec:monadofadjoint}:
\begin{lstlisting}{coq}
Definition Mgcm := Monad_of_adjoint Agcm.
\end{lstlisting}

The very last step is to use the function \L!Monad_of_category_monad!
of Sect.~\ref{sec:fun_mon} to recover a monad compatible with the
\monae{} formal framework of monadic equational reasoning\footnote{We
  can also recover the probability monad of~\cite{affeldt2019mpc}
  which is definitionally equal to {\tt
    Monad\us{}of\us{}category\us{}monad (Monad\us{}of\us{}adjoint
    (adj\us{}comp AC A0))}.}:
\begin{lstlisting}{coq}
Definition gcm := Monad_of_category_monad Mgcm.
\end{lstlisting}

\subsubsection{Informal Description of the Join of the Monad}

At this stage, it is worth taking a step back to check that the
join of the monad we have built indeed corresponds to the
intuition one can have of the execution of a program mixing
probabilistic choice and nondeterministic choice.
Provided we ignore the function $\varepsilon_C$ (the counit of the
adjunction $F_C \dashv U_C$, which, as we already explained in
Sect.~\ref{sec:fcuc}, is here essentially for technical reasons), the
join operator can informally be explained as the following function:
$$
\varepsilon_1 \circ (\lambda\,X.\, \image{\varepsilon_0}{X}).
$$
The input of this function is indeed %
\L!necset {dist (necset {dist T})}!, i.e., it takes non-empty sets of
distributions.
The function $\varepsilon_0$ (Sect.~\ref{sec:counitunit0}) computes
barycenters, so that when applied the right-hand side of the function
composition returns an object of type \L!necset (necset {dist T})!.
The function $\varepsilon_1$ (Sect.~\ref{sec:counitunit1}) computes
the hull of the union of its input, which results in an object of
type~\L!necset {dist T}!, as expected.

\section{The Properties of Combined Choice of the \GCM{}}
\label{sec:prop_comb_choice}

The very last step of our construction is to show that the \gcm{}
(that we obtained as a result of the previous
section---Sect.~\ref{sec:adjoint_functors}) satisfies the expected
distributivity axioms that we discussed in Sect.~\ref{sec:target} {\em
  and\/} to check that it is meaningful, i.e., that it really
distinguishes the different choice operators.
This corresponds to~\cite[file \L!altprob_model.v!]{monae} in the
accompanying development.

\subsection{The \GCM{} has the Properties of Combined Choice}

First, we start by defining nondeterministic choice for the \gcm{}
using a binary version of the operator $\lubop$ of Sect.~\ref{sec:scsl}:
\begin{lstlisting}{coq}
Definition alt A (x y : gcm A) : gcm A := x [+] y.
\end{lstlisting}
We construct a monad \L!gcmA! implementing \L!altMonad! by proving the
following properties, which are essentially consequences of the
properties of the operator $\lubop$:
\begin{lstlisting}{coq}
Lemma altA A : associative (@alt A).
Lemma bindaltDl : BindLaws.left_distributive (@monad.Bind gcm) alt.
Definition gcmA : altMonad := MonadAlt.Pack ...
\end{lstlisting}
We extend the monad \L!gcmA! to the monad \L!gcmACI! that implements
\L!altCIMonad!:
\begin{lstlisting}{coq}
Lemma altxx A : idempotent (@Alt gcmA A).
Lemma altC A : commutative (@Alt gcmA A).
Definition gcmACI : altCIMonad := MonadAltCI.Pack ...
\end{lstlisting}

Second, we go on defining probabilistic choice for the \gcm{}
using the operator of convex spaces:
\begin{lstlisting}{coq}
Definition choice p A (x y : gcm A) : gcm A := x <| p |> y.
\end{lstlisting}
Most properties are direct consequences of the properties of convex
spaces, and they lead to the definition of the monad \L!gcmp!
that implements \L!probMonad!:
\begin{lstlisting}{coq}
Lemma choice0 A (x y : gcm A) : x <| 0%:pr |> y = y.
Lemma choice1 A (x y : gcm A) : x <| 1%:pr |> y = x.
Lemma choiceC A p (x y : gcm A) : x <|p|> y = y <|p.~%:pr|> x.
Lemma choicemm A p : idempotent (@choice p A).
Lemma choiceA A (p q r s : prob) (x y z : gcm A) :
  p = (r * s) :> R /\ s.~ = (p.~ * q.~)%R ->
  x <| p |> (y <| q |> z) = (x <| r |> y) <| s |> z.
Definition gcmp : probMonad := MonadProb.Pack ...
\end{lstlisting}

Finally, we prove left-distributivity of bind over the probabilistic
choice and right-distributivity of the probabilistic choice over the
nondeterministic choice
\begin{lstlisting}{coq}
Lemma bindchoiceDl p : BindLaws.left_distributive (@monad.Bind gcm) (@choice p)
Lemma choicealtDr A (p : prob) :
  right_distributive (fun x y : gcmACI A => x <| p |> y) Alt.
\end{lstlisting}
and use these lemmas to instantiate \L!atlProbMonad! into the monad
\L!gcmAP!:
\begin{lstlisting}{coq}
Definition gcmAP : altProbMonad := MonadAltProb.Pack ...
\end{lstlisting}
This completes the construction of the monad proposed by Gibbons et
al.~\cite{gibbons2011icfp,abousaleh2016}.

\subsection{The Combined Choice is not a Trivial Theory}
\label{sec:nocollapse}

We conclude this section with a formal check that probabilistic choice
in our axiom system of combined choice is not trivial, meaning that it indeed
distinguishes different probabilities.
It is sufficient to check that there exists a model which is not trivial in this
sense, and our construction of \gcm{} serves this purpose nicely:
\begin{lstlisting}{coq}
Example gcmAP_choice_nontrivial (p q : prob) :
  p <> q ->
  Ret true <|p|> Ret false <> Ret true <|q|> Ret false :> gcmAP bool.
Proof.
...
Qed.
\end{lstlisting}
Here \L!:> gcmAP bool! indicates the type of this inequality, which
forces the resolution of monadic operations inside our instance of
\L!altProbMonad!.  The proof just requires to unfold definitions and
provides further evidence that the \gcm{} is not a trivial model.

\section{Application: Mechanization of the Monty Hall Problem}
\label{sec:monty_hall}

As an application of \L!altProbMonad!, we provide a mechanization of
the Monty Hall problem using probability {\em and\/} nondeterminism as
described by Gibbons~\cite[Sect.~6.1]{gibbons12utp} (we have also
mechanized a purely probabilistic
variant~\cite[Sect.~6]{gibbons12utp}\cite[Sect.~8.1]{gibbons2011icfp}
as well as a forgetful variant~\cite[Sect.~7.2]{gibbons12utp}).

Let us recall the Monty Hall problem. The player is given a choice of
three doors: there is a car behind one door and there are goats behind
the other doors. First, the player picks one door and the host opens
one of the other doors behind which there is a goat. The player is
then asked whether he/she wants to stick to his/her first choice or
switch to the other door. It turns out that the best strategy is to
switch, even though this appears to be counterintuitive for many, as
shown by the controversy the problem sparked when first exposed in the
specialized press.

\subsection{Problem Setting}

Let us consider the datatype \L!door! consisting of three different
doors \L!A!, \L!B!, and \L!C! (\L!doors! is a list consisting of these three doors).
The host hides the car behind one of the three doors chosen nondeterministically
(hence \L!altMonad!) (below, \L!def! has type \L!door! unless quantified):
\begin{lstlisting}{coq}
Definition hide_n {M : altMonad} : M door := arbitrary def doors.
\end{lstlisting}
The function \L!arbitrary! takes a default element and a list and
returns an element of the list chosen nondeterministically (or the
default element if the list is empty). It is defined using standard
functions as follows:
\begin{lstlisting}{coq}
Definition arbitrary {M : altMonad} {A : Type} (def : A) : seq A -> M A :=
  foldr1 (Ret def) (fun x y => x [~] y) \o map Ret.
\end{lstlisting}

The player picks one of the doors uniformly at random
(using \L!probMonad!):
\begin{lstlisting}{coq}
Definition pick {M : probMonad} : M door := uniform def doors.
\end{lstlisting}
The function \L!uniform! is defined using the binary probabilistic
choice as follows:
\begin{lstlisting}{coq}
Fixpoint uniform {M : probMonad} {A : Type} (def : A) (s : seq A) : M A :=
  match s with
    | [::] => Ret def
    | [:: x] => Ret x
    | x :: xs =>
      Ret x <| (/ IZR (Z_of_nat (size (x :: xs))))%:pr |> uniform def xs
  end.
\end{lstlisting}

The host teases the player by opening a door, which is nor the one
hiding the car neither the one picked by the player, chosen
nondeterministically:
\begin{lstlisting}{coq}
Definition tease_n {M : altMonad} (h p : door) : M door :=
  arbitrary def (doors \\ [:: h; p]).
\end{lstlisting}

We can now arrange above elements chronologically to represent a game,
the latter being parameterized by the strategy of the player:
\begin{lstlisting}{coq}
(* generic game *)
Definition monty {M : monad} hide pick tease
    (strategy : door -> door -> M door) :=
  do h <- hide ;
  do p <- pick ;
  do t <- tease h p ;
  do s <- strategy p t ;
  Ret (s == h).

(* nondeterministic variant *)
Variable M : altProbMonad.
Definition play_n (strategy : door -> door -> M door) : M bool :=
  monty hide_n (pick def) tease_n strategy.
\end{lstlisting}

We finally provide the two possible strategies.
The ``stick'' strategy is defined by returning the already-chosen door:
\begin{lstlisting}{coq}
Definition stick {M : monad} (p t : door) : M door := Ret p.
\end{lstlisting}
The ``switch'' strategy is defined by returning the other door (the
one that was nor picked neither used for teasing):
\begin{lstlisting}{coq}
Definition switch {M : monad} (p t : door) : M door :=
  Ret (head def (doors \\ [:: p ; t])).
\end{lstlisting}

\subsection{Switch is Better than Stick}

One can prove that the ``switch'' strategy is better than the ``stick''
strategy by comparison with a biased coin, defined as follows
(definiton from Sect.~\ref{sec:target} reproduced here for the
convenience of the reader):
\begin{lstlisting}{coq}
Definition bcoin {M : probMonad} (p : prob) : M bool :=
  Ret true <| p |> Ret false.
\end{lstlisting}

More precisely, one can show that the ``switch'' strategy is as good
as a $2/3$-biased coin (recall from Sect.~\ref{sec:target} that \L!(/ 3).~%:pr!
is the probability $1 - 1/3 = 2/3$):
\begin{lstlisting}{coq}
Lemma monty_switch : play_n (switch def) = bcoin (/ 3).~%:pr.
\end{lstlisting}
The proof goes as follows.
\begin{enumerate}
\item The left-hand side \L+play_n (switch def)+ can be rewritten as:
\begin{lstlisting}{coq}
hide_n >>= (fun h => pick def >>= (fun p => tease_n h p >>=
    (fun t => Ret (h == head def (doors \\ [:: p; t])))))
\end{lstlisting}
This step essentially amounts to use the property that the unit is the
left neutral of bind.
\item The rightmost continuation can furthermore be rewritten to lead to:
\begin{lstlisting}{coq}
hide_n >>= (fun h => pick def >>= (fun p => tease_n h p >>
    (if h == p then Ret false else Ret true)))
\end{lstlisting}
This step is essentially by case analysis on \L!h == p! and observation of the expression
\begin{lstlisting}{coq}
head def (doors \\ [:: p; t]).
\end{lstlisting}
When \L!h == p!, this expression cannot be \L!h!.
When \L+h != p+, it is \L!h!.
\item Since teasing does not influence the outcome anymore, the left-hand side can 
furthermore be simplified into:
\begin{lstlisting}{coq}
hide_n >>= (fun h => pick def >>= (fun p => Ret (h != p)))
\end{lstlisting}
The main lemma needed for this step can be stated in a generic way as follows:
\begin{lstlisting}{coq}
Lemma arbitrary_inde (M : altCIMonad) T (a : T) s U (m : M U) :
  0 < size s -> arbitrary a s >> m = m.
\end{lstlisting}
\item The last step produces the expected biased coin
\L!bcoin (/ 3).~%:pr.!
This is captured by the following lemma:
\begin{lstlisting}{coq}
Lemma bcoin23E :
  arbitrary def doors >>=
    (fun h => uniform def doors >>= (fun p => Ret (h != p))) =
  bcoin (/ 3).~%:pr.
\end{lstlisting}
Its proof essentially appeals to the properties of probabilistic
choice as specified by the interface of \L!probMonad! seen in
Sect.~\ref{sec:target} and to the fact that bind left-distributes over
nondeterministic choice, a property of \L!altMonad!.
\end{enumerate}

On the other hand, the ``stick'' strategy is as good as a $1/3$-biased coin:
\begin{lstlisting}{coq}
Lemma monty_stick : play_n stick = bcoin (/ 3)%:pr.
\end{lstlisting}
The proof is a bit simpler. It suffices to observe that the teasing
does not influence the outcome and use the lemma \L!arbitrary_inde!.
It is completed by computations similar to the last step of the proof
for the ``switch'' strategy and uses the fact that bind
left-distributes over nondeterministic choice and probabilistic
choice.

As the reader has observed in this section, the example of the Monty
Hall problem uses only the interfaces of the involved monads,
including the \L!altProbMonad!, and we know for sure that its
interface is correct since we have a formal model since
Sect.~\ref{sec:formal_gcm}.

\section{Related Work}
\label{sec:related}

We have already commented on several related work throughout this
paper. We add in this section further comments that are better
explained now that we have completed the technical presentation of our
contributions.

The formalization of convex spaces comes from~\cite{saikawa2020cicm}.
This work develops applications of convex spaces such as convex and
concave functions and formalizes equivalences between various
axiomatizations of binary and multiary convex operators. Here, we use
the multiary convex combination operator in Sect.~\ref{sec:f0u0}, we
further develop the theory of affine functions, and we extend convex
spaces to build the convex powerset functor.

In our formalization of semicomplete semilattices (in
Sect.~\ref{sec:semicompsemilatt}), the nondeterministic choice is
modeled as an infinitary operator.
This is similar to Beaulieu's ``infinite nondeterministic
choice''~\cite[Def.~3.2.3]{beaulieu2008phd} and, at first sight,
looks different from Cheung's approach, who models nondeterministic
choice as a binary operator~\cite[Sec.~6.3.1]{cheungPhD2017}.
In Sect.~\ref{sec:counitunit1}, we explained that Cheung also
implicitly uses an infinitary version of his operator and that we find
an infinitary operator to be more comfortable and clearer from the
viewpoint of formalization.

The monad for probability and nondeterminism can also be presented
using finitely-generated convex sets of
distributions~\cite[Sect.~3.1]{bonchi2019lics}.
Here, we did not insist on having finitely-generated convex sets
because our first attempt at doing so led to technically involved
formal proofs.
Now that we have completed our formalization, it should be easier to
extend it with finitely-generated convex sets.
Indeed, looking at \cite{bonchi2020theory}, we recognize several
technical results that we have already formalized (e.g., parts of
Lemma~4.4).
Concretely, the approach would start by defining the data structure
for the non-empty finitely-generated convex sets by adding an axiom
for the existence of a finite generator to the type \L!necset!, and
then by replaying and fixing the proofs (the category part of our
framework should stay unchanged).
This could open the door to the construction of an executable model.
%
%Alternatively, \cite{bonchi2020presenting} suggests that we could
%work only with generators but the implications of this approach for
%the formalization work are difficult to evaluate beforehand.

The \gcm{} is not the first example of a formalized monad that
combines probabilistic and nondeterministic choices: Tassarotti and
Harper~\cite{tassarotti2019popl} already formalized in \coq{} the
indexed valuation monad by Varacca and Winskel that we already
mentioned in Sect.~\ref{sec:alternative}. In this monad, probabilistic
is not idempotent and therefore it is not suitable for our
purpose. Our formalization looks arguably more modular than the one by
Tassarotti and Harper who build their monad in a direct manner.

We have been formalizing one model that combines probabilistic and
nondeterministic choices: the one advocated by Gibbons et al.\ because
it fits well with functional programming.
%x[
pGCL~\cite{mciver2005} is another such model that has been formalized
in the Isabelle/HOL proof assistant~\cite{cock2014} (as such it
qualifies as the first formalized model that provides both
probabilistic and nondeterministic choices).
However, its default semantics is given in different terms (using
predicate transformers, no category theory involved, refinement
instead of equations) so that the formalizations of the geometrically
convex monad and of pGCL turn out to be different tasks.
McIver and Morgan's book~\cite{mciver2005} contains also another
semantics, the relational demonic semantics whose mathematical
construction (Definition~5.4.4) is similar to Cheung's.  Yet, it is
not presented as a monad with algebraic laws, which is a crucial
aspect of our framework, and it has not been formalized.

There is a number of formalizations of category theory in proof assistants (many
of which being listed by Gross et al.~\cite{gross2014itp}).  However, we could
not find a readily usable formalization of concrete categories in \coq. For
example, UniMath is a large \coq{} library that aims at formalizing mathematics
using the univalent point of view~\cite{UniMath}. It contains a substantial
formalization of abstract categories but does not seem to feature a
formalization of concrete categories.
Since we needed only a handful of theorems about category theory, we
formalized concrete categories from scratch and developed their
theories as a generalization of \monae{} (in
Sect.~\ref{sec:categories}).

The idea of using categories as a package to handle functions with proofs was
already presented by McBride~\cite[Chapter~7, Section~3.1]{mcbride1999phd}.
He also proposed the use of concrete categories for such a lightweight use of
category theory, noting that the convertibility of terms is an easier way than
propositional equality to handle the equational laws for morphisms, such as unit
and associativity laws.
His formal definition of categories differs from ours in that it is also
indexing on hom-sets, while in our definition, hom-sets are embedded as
predicates.  This difference further affects later definitions
such as functors.
Our definition makes it clearer that concrete categories are shallow
embeddings of categories.

\section{Conclusion and Future Work}
\label{sec:conclusion}

In this paper, we proposed a formalization in the \coq{} proof
assistant of an infinitary version of the \gcm{}, a monad that
combines probabilistic and nondeterministic choice with idempotence of
probabilistic choice. To the best of our knowledge, this is the first
formalization of such a monad.
Our development led us to develop several formal mathematical
theories of broader interest such as a formalization of the convex
powerset functor and a formalization of concrete categories.
A direct application was to complete an existing formalization of
monadic equational reasoning which was lacking the model of the
combined interface of probabilistic and nondeterministic choices
and which we illustrated with an extended example.
We could also use our model to check that the probabilistic
operator does not collapse with Gibbons et al.'s choice of axioms.

We formalized an infinitary nondeterministic choice operator.
As we discussed in Sect.~\ref{sec:related}, it would be interesting to
formalize a finitary one with the insights from recent work on
finitely-generated convex sets~\cite{bonchi2020presenting}.
Our experiment is an example of combination of two monads that requires a
substantial amount of work. There also exist a number of generic results about
the combination of monads such as distributive laws~\cite{zwart2018arxiv} or
weak ones~\cite{goy2020lics} that would deserve formalization.
By introducing a formalization of concrete categories to support the
construction of the \gcm{}, our work also raises the question of the
generalization of \monae{}~\cite{affeldt2019mpc} from its specialization
to the {\bf Set} category.

\paragraph*{Acknowledgments}~%
We acknowledge the support of the JSPS KAKENHI Grant Number 18H03204
and the JSPS-CNRS bilateral program ``FoRmal tools for IoT
sEcurity'' (PRC2199), and thank all the participants of these
projects for fruitful discussions. We also thank Cyril Cohen and
Shinya Katsumata for guidance about the formalization of monads,
Kazunari Tanaka who contributed to the formalization of categories,
Jeremy Gibbons and Joseph Tassarotti for their comments.

\bibliographystyle{apalike}
\bibliography{altprob_bib}

\end{document}